\documentclass[journal,draftclsnofoot,onecolumn,12pt]{IEEEtran}
\usepackage{amsmath}
\usepackage{amssymb}
\usepackage{amsthm}
\usepackage{mathrsfs}
\usepackage{bm}
\usepackage{graphicx}
\usepackage{color}
\usepackage{epsfig}

\title{Multicasting in Large Wireless Networks: Bounds on the Minimum Energy per Bit}
\author{Aman~Jain,~Sanjeev~R.~Kulkarni,~and~Sergio~Verd\'u}
\date{\today}

\begin{document}

\maketitle
\newtheorem{theorem}{Theorem}
\newtheorem{lemma}{Lemma}
\newtheorem{corollary}{Corollary}

\renewcommand{\thefootnote}{\fnsymbol{footnote}}
\setcounter{footnote}{2}

\newcommand{\whp}{\emph{whp}}
\newcommand{\R}{\mathbb{R}}
\newcommand{\RP}{\mathbb{R^+}}
\newcommand{\C}{\mathbb{C}}
\newcommand{\N}{\mathbb{N}}
\renewcommand{\P}{P}
\newcommand{\E}{\mathbb{E}}

\newcommand{\rk}{r}
\newcommand{\PMk}{P_M(k)}
\newcommand{\PFk}{P_F(k)}

\newcommand{\rec}{r}
\newcommand{\Rset}{\mathcal{R}}

\newcommand{\A}{\ensuremath{\mathsf{A}}}
\newcommand{\G}{\ensuremath{\mathsf{G}}}
\newcommand{\Ak}{\ensuremath{A_k}}
\newcommand{\ak}{\ensuremath{a}}
\newcommand{\akk}{\ensuremath{a_k}}
\newcommand{\tauk}{\ensuremath{\tau}}
\newcommand{\EbNo}{\ensuremath{\left( \frac{E_b}{N_0} \right)}}
\newcommand{\EbNomin}{\ensuremath{ {\frac{E_b}{N_0}}_{\text{min}}}}
\newcommand{\Flood}{\texttt{FLOOD}$\left( {E_b}_1, {E_b}_2 \right)$ }
\newcommand{\mkgood}{\ensuremath{s^{(k)}}}
\newcommand{\mkneigh}{\ensuremath{s^{(k)}}}
\newcommand{\mk}{\ensuremath{s^{(k)}}}
\renewcommand{\l}{\ensuremath{L}}
\newcommand{\Et}{\ensuremath{E_{\text{total}}}}
\newcommand{\Ik}{\ensuremath{I_k}}
\newcommand{\Ikk}{\ensuremath{{I_k}'}}
\newcommand{\bigdiv}{\mbox{\huge $/$ }}
\newcommand{\cell}{C}
\renewcommand{\k}{\ensuremath{\nu}}

\begin{abstract}

We consider scaling laws for maximal energy efficiency of communicating a message to all the nodes in a wireless network, as the number of nodes in the network becomes large. Two cases of large wireless networks are studied --- dense random networks and constant density (extended) random networks. In addition, we also study finite size regular networks in order to understand how regularity in node placement affects energy consumption.

We first establish an information-theoretic lower bound on the minimum energy per bit for multicasting in arbitrary wireless networks when the channel state information is not available at the transmitters. Upper bounds are obtained by constructing a simple flooding scheme that requires no information at the receivers about the channel states or the locations and identities of the nodes. The gap between the upper and lower bounds is only a constant factor for dense random networks and regular networks, and differs by a poly-logarithmic factor for extended random networks. Furthermore, we show that the proposed upper and lower bounds for random networks hold almost surely in the node locations as the number of nodes approaches infinity.

\end{abstract}

%\begin{IEEEkeywords}
%Cooperative communication, minimum energy per bit, multicasting, wideband communication, wireless networks.
%\end{IEEEkeywords}

\section{Introduction}

\subsection{Prior Work}
\label{priorwork}
Determining the energy efficiency of a point-to-point channel is a fundamental information-theoretic problem. While the minimum energy per bit requirement for reliable communication is known for a general class of channels \cite{UnitCost,Spectral}, the problem is considerably more complicated for networks. Even when just one helper (\emph{relay}) node is added to the two terminal AWGN channel, the minimum energy per bit is still unknown, though progress has been made in \cite{Zahedi}, \cite{YCG}, \cite{CYG}. The minimum energy per bit for the general Gaussian multiple-access channel, the broadcast channel and the interference channel has been considered in \cite{UnitCost,Spectral,Lapidoth,Caire}. As the number of relays $k$ in a network grows, one can ask whether the energy efficiency improves, and by what rate. It is shown in \cite{Dana} that a two-hop \emph{distributed beamforming} scheme is energy efficient for dense random networks, with the energy requirement falling as $\Theta ({1}/{\sqrt{k}})$. In this scheme, however, the relay nodes require information of the channel states of the forward and backward links. It is not clear how to extend the same idea to noncoherent or to multicasting scenarios. See \cite{Mammen} and references therein, for the energy efficiency of multi-hopping in a unicast random network setting. 

Cooperation between nodes (also known as \emph{cooperative diversity}) leads to capacity or reliability gains even with simple communication schemes (e.g., see \cite{Wornell,Sendonaris,Sendonaris2} amongst others). One of the simple ideas for cooperation in a multicast setting is to let several nodes transmit the same signal (at lower power levels), so that each receiving node can combine several low reliability signals to construct progressively better estimates. This scheme works only if all the nodes retransmit the same message. The works of \cite{Maric}, \cite{ScaglioneOLA1} and \cite{ScaglioneOLA2} present such multi-stage decode and forward schemes to reduce the transmission energy in a network. The question about the best scheme of this nature can be formulated as an \emph{Optimal Cooperative Broadcast} \cite{ScaglioneOLA2,ScaglioneJSAC} or an \emph{accumulative broadcast} problem \cite{Maric}. In these formulations, first, an optimal transmission order for the nodes is constructed. Given such an order, the transmission energy is then minimized by solving a linear program for the power distribution, subject to the condition that all the nodes receive a minimum amount of power. 

The problem of communicating the same message to a set of nodes (multicasting) in a network with minimum energy consumption, has drawn a lot of research interest. For the case of wired networks, the problem can be formulated as the well known \emph{Minimum cost Spanning Tree} problem. However, for wireless networks, there is an inherent \emph{wireless multicast advantage} \cite{BT} that allows all the nodes within the coverage range to receive the message at no additional cost. The minimum energy broadcast problem for the wireless networks was formulated as a \emph{broadcast tree problem} in \cite{BT}. The formulation based on wireless multicast advantage, however, still misses the advantage of overhearing other transmissions over the network. This advantage is important in the setting where the same message is passed around the network. Such an advantage has been referred to as \emph{Cooperative Wireless Advantage} (CWA) in \cite{ScaglioneOLA2}. A more fundamental approach to the modelling and analysis of wireless networks may yield better results based on exploiting the broadcast nature of wireless communications. 

The power efficiency of a decode and forward multicasting scheme for dense random networks has been studied in \cite{ScaglioneJSAC}. Their work focuses on minimizing the power requirement in a non-zero power and finite bandwidth regime, under a different system model. Moreover, in \cite{ScaglioneJSAC}, achievability schemes were presented for dense networks, whereas our interest is in the order of growth of energy requirement for simpler power allocation (uniform) for both dense and extended networks. A major difference in our setup from the previous works is our emphasis on minimal network and channel state information. This implies, among other things, that no centrally optimized transmission or power policies can be implemented.

Scaling laws for the upper and lower bounds on the multicast capacity are considered in \cite{MergenGastpar1} and \cite{MergenGastpar2} for the dense and extended network cases respectively. The question of multicast capacity for multihopping is addressed in \cite{Zheng} and \cite{Riedi}. Energy-efficient area coverage using a multi-stage decode and forward scheme is studied in \cite{ScaglioneITTrans}.

\subsection{Summary of Results}
 \label{summary}

In this work, our aim is to determine the maximum possible energy efficiency (i.e., minimum transmission energy per information bit) for multicasting in various wireless networks when there is no constraint on the bandwidth. Our focus is on the particular multicasting setting where all the nodes are interested in a common message. Besides developing suitable converse bounds, we also show how cooperative communication is instrumental to approach them.
  
We first present, in Theorem \ref{thm1}, an information-theoretic lower bound on the energy requirement for multicasting in arbitrary wireless networks. The lower bound is shown to be inversely proportional to the \emph{effective radius} of the network, which is a fundamental property of the network and is determined by the gains between the nodes and the set of destination nodes. This bound is applicable whenever channel state information is not available at the transmitters. 

For the achievability part, we propose a simple flooding algorithm that does not require knowledge of the node locations, identities or channel states. For the networks that we consider, we show that just the information about the number of nodes, the area of the network and the fading statistics is sufficient to achieve the same order of energy scaling as that of the schemes with considerably more knowledge.

The converse and achievability bounds on the minimum energy requirement per information bit are then evaluated for two cases of large random networks and for finite regular networks. Since the problem of studying the energy efficiency of multicasting in general wireless networks is non-trivial, following recent trends, we instead focus on the order of scaling of the energy efficiency for large random networks. Such scaling laws reveal the major factors affecting the energy efficiency when the number of nodes $k$ is large. 

The physical channel is modeled as a fading channel subject to Gaussian noise. We operate in the wideband regime, which is essential to maximize the energy efficiency in a point-to-point Gaussian channel \cite{Spectral}. Since there is no bandwidth constraint, we allot a separate wide band transmission channel to each node. The power-constrained, wideband multicast setting considered here is particularly relevant to sensor networks \cite{Sensor}.

The gain between a pair of nodes is determined by the distance between the nodes according to a \emph{path loss model}. Specifically, we model the power gain between any two nodes as falling off as ${r^{-\alpha}}$ with the distance $r$ between the nodes, for $\alpha>2$. Furthermore, we assume that the gain never exceeds $\bar{g}$ for any distance between the nodes.

The different kinds of networks that we study here are:
\begin{itemize}
\item \emph{Large dense random networks}, where the $k-1$ non-source nodes are placed randomly i.i.d. uniformly over a square area of size $\Ak$ which increases as $o\left( k/\log k \right)$.
\item \emph{Large extended random networks}, where the $k-1$ non-source nodes are placed randomly i.i.d. uniformly over a square area of size $\Ak$ which increases linearly with $k$. 
\item \emph{Finite regular networks}, where the network is divided into small square cells and each cell is assumed to contain exactly one node. Furthermore, the nodes are confined to a certain fraction of area within these cells. 
\end{itemize}
For the case of large networks, we are interested in the asymptotic analysis (as $k\rightarrow \infty$) of the upper and lower bounds on the energy requirement per bit. On the other hand, regular networks are studied for all values of $k\geq 2$.

There has been considerable research into multicasting algorithms by the networks community (see, e.g., \cite{flood,Camp} and references therein). We borrow one such simple technique --- \emph{flooding} \cite{flood,Obraczka} based on repetition-coding, to achieve our goals. The central idea is to collect energy from multiple transmissions to reconstruct the original message  \cite{Maric,ScaglioneJSAC}. 

The rest of our paper is structured as follows. In Section \ref{notation}, we introduce the system model. In Section \ref{lowerbound}, we prove a general result about the minimum energy requirement of multicasting in a wireless network. A form of flooding algorithm is introduced in Section \ref{algoflood}. 
In Section \ref{denserandomnetwork}, dense random networks are introduced and their minimum energy per bit is shown to scale linearly with area. In Section \ref{randomextendednetwork}, extended random networks are studied. In this case, the minimum energy per bit is lower bounded as $\Omega (k)$, with the constant depending on the node density. Our flooding algorithm is shown to come within a poly-logarithmic (in $k$) factor of the lower bound. For both cases of large random networks, the bounds hold almost surely in the placement of nodes as $k \rightarrow \infty$. 
In Section \ref{regularnetwork}, we take up the case of finite regular networks. In general, the lower bound on minimum energy per bit of a regular network can depend on both the number of nodes and the node density. However, the energy consumption of flooding algorithm is always within a constant factor of the lower bound.

\section{System Model}
\label{notation}

\subsection{Channel Model}

We deal with a discrete-time complex additive Gaussian noise channel with fading. Suppose that there are $k$ nodes in the network, with node 1 being the source node. Let the node $i\in \{1,...,k\}$ transmit $x_{i,t}\in \R$ at time $t$, and let $y_{j,t}\in \C$ be the received signal at any other node $j\in \{1,...,i-1,i+1,...,k\}$. The relation between $x_{i,t}$ and $y_{j,t}$ at any time $t$, is given by
\begin{equation}
y_{j,t}=\sum_{i=1}^{k}h_{ij,t}x_{i,t}+z_{j,t} \label{channeleq}
\end{equation}
where $z_{j,t}$ is circularly symmetric complex additive Gaussian noise at the receiver $j$, distributed according to $\mathcal{CN}(0,N_0)$. The noise terms are independent for different receivers as well as for different times. 
The fading between any two distinct nodes $i$ and $j$ is modeled by complex-valued circularly symmetric random variables $h_{ij,t}$ which are i.i.d. for different times. We assume that $h_{ii,t}=0$ for all nodes $i$ and times $t$.
Also, for all $(i,j)\neq (l,m)$, the pair $h_{ij,t}$ and $h_{lm,t}$ is independent for all time $t$. 
\emph{Absence of channel state information at a transmitter $i$} implies that  $x_{i,t}$ is independent of the channel state realization vector $(h_{i1,t},h_{i2,t},...,h_{ik,t})$ from node $i$ to all other nodes, for all times $t$. The quantity $\E[|h_{ij}|^2]$ is referred to as the \emph{channel gain} between nodes $i$ and $j$.

\subsection{Problem Setup}

All the nodes in the network are identical and are assumed to have receiving, processing and transmitting capabilities. The nodes can also act as relays to help out with the task of communicating a message to the whole network. The total energy consumption of the network is simply the sum of transmission energies at all the nodes. To define a multicast relay network, we extend the three terminal relay channel setting of \cite{CoverGamalRelay} to include multiple relays and multiple destination nodes. An error is said to have occurred when any of the intended nodes fails to decode the correct message transmitted by the source.

Consider a code for the network with block length $n\in \mathbb{N}$. For $i=1,...,k$, the codeword at node $i$ is $n$ symbols long, denoted by $x_{i}^{(n)}=(x_{i,1},x_{i,2},...,x_{i,n})\in \mathbb{C}^n$. If the message set at the source node (node 1) is $\mathcal{M}=\{1,2,...,M\}$, then the codeword $x_1^{(n)}(m)$ is determined by the message $m$ chosen equiprobably from the message set. At any other node $i\in \{2,...,k\}$, the codeword $x_i^{(n)}$ is a function of the channel outputs $y_i^{(n)}=(y_{i,1},y_{i,2},...,y_{i,n})$ at the node. Due to causality, the $t^{\text{th}}$ symbol $x_{i,t}$ of $x_i^{(n)}$ is a function of the first $t-1$ inputs at the node, i.e., $x_{i,t}=x_{i,t}(y_i^{(t-1)})$. This function, which defines the input-output relation at a relay, is also called the \emph{relay function}.

At each non-source node $i$, in addition to a relay function, there may also be a \emph{decoding function} (depending on whether the node is a \emph{destination node}) which decodes a message $\hat{m}_i\in \mathcal{M}$ based on the $n$ channel outputs $y_{i}^{(n)}$ at the node. Therefore, $\hat{m}_i=\hat{m}_i(y_{i}^{(n)})$. 

Suppose that only a subset $\Rset\subseteq \{2,...,k\}$ (also called the \emph{destination set}) of the nodes is interested in receiving the message from the source node. When $\Rset$ contains two or more nodes, it is called a \emph{multicast} setting. 

The probability of error of the code is defined as
\begin{equation}
{P}_e \triangleq \frac{1}{M}\sum_{m\in \mathcal{M}} P_e[m] 
\end{equation}
where, 
\begin{equation}
P_e[m]\triangleq \P[\exists i\in \Rset : \hat{m}_i \neq m | m \text{ is the message}] \label{equationpoe}
\end{equation}
Note that the error event at a single node is a subset of the error event defined above. Clearly, $P_e$ is at least as big as the probability of error at any subset of the nodes in $\Rset$. 

Next, we define the energy per bit of the code. Let $E_{\text{total}}$ be the expected total energy expenditure (for all nodes) of the code, i.e.,
\begin{equation}
E_{\text{total}} \triangleq \E\left[ \sum_{i=1}^k \sum_{t=1}^n |x_{i,t}|^2  \right]
\end{equation}

The \emph{energy per bit} of the code is defined to be
\begin{equation}
E_b \triangleq \frac{E_{\text{total}}}{\log_2 M} \label{energyperbit}
\end{equation}

Let 
\begin{equation}
E_{i,t} \triangleq \E[|x_{i,t}|^2]
\end{equation}
be the expected energy spent transmitting the $t^{\text{th}}$ symbol at node $i$. Then, the energy per bit of the code can also be written as
\begin{align}
E_b & = \frac{1}{\log_2 M} \sum_{i=1}^{k} \sum_{t=1}^n E_{i,t}
\end{align}
Note that, in each case, the expectation is over the message, noise and fading.

An $(n,M,E_{\text{total}},\epsilon)$ code is a code over $n$ channel uses, with $M$ messages at the source node, expected total energy consumption at most $E_{\text{total}}$ and a probability of error at most $0\leq \epsilon<1$.

In \cite{UnitCost}, \emph{channel capacity per unit cost} was defined for a channel without restrictions on the number of channel uses. Here, we are interested in the reciprocal of this quantity.

\emph{Definition:} Given $0\leq \epsilon <1$, $E_b \in \mathbb{R}_+$ is an $\epsilon$-achievable energy per bit if for every $\delta>0$, there exists an $E_0\in \mathbb{R}_+$ such that for every $E_{\text{total}}\geq E_0$ an $(n,M,E_{\text{total}},\epsilon)$ code can be found such that
\begin{equation}
\frac{E_{\text{total}}}{\log_2 M} < E_b + \delta
\end{equation}
$E_b$ is an achievable energy per bit if it is $\epsilon$-achievable energy per bit for all $0<\epsilon<1$, and the \emph{minimum energy per bit} ${E_b}_{\min}$ is the infimum of all the achievable energy per bit values. Sometimes, we deal with the normalized (w.r.t. noise spectral density $N_0$) version of ${E_b}_{\min}$, which is represented by $\EbNomin$.

In the power constrained regime of wireless networks, ${E_b}_{\min}$ is a sensible measure of how many bits of information can be reliably transmitted for a given (large enough) energy quota. Alternately, energy per bit could also be defined as the energy required per bit for all (large enough) sizes of the message set. All the results given in this paper hold under this definition as well.

\emph{Minimal information framework:} We derive scaling results for minimum energy per bit for different classes of networks. For a given number of nodes $k$, each class of networks has a set of possible network realizations. Our aim is to achieve low energy consumption per bit using no information at the nodes about the actual network realization (i.e., node locations). In addition, we also assume that the nodes have no information about the channel states. All the non-source nodes have the same relay and decoding functions.

Providing local or global information to the nodes enlarges the set of possible coding schemes. Our converse results allow coding schemes to rely on any such information except for channel state information at the transmitters. 

\section{A Lower Bound on the Minimum Energy Per Bit}
\label{lowerbound}

In this section, in Theorem \ref{thm1}, we show an information theoretic lower bound on the minimum energy per bit for multicast in an arbitrary network. The bound depends on the destination nodes and the channel gains, through \emph{effective network radius} defined below. It holds for any communication scheme where channel states are not known at the transmitters.

\begin{theorem}
\label{thm1}
In a network with $k$ nodes, where node 1 is the source node and the destination set is $\Rset \subset \{2,...,k\}$, the required minimum energy per bit satisfies
\begin{equation}
\EbNomin \Bigl( \Rset \Bigr) \geq \frac{\log_e2}{G (\Rset)} \label{thm1statement1}
\end{equation}
where $G$ is the \emph{effective network radius} defined as
\begin{equation}
G (\Rset) \triangleq \frac{1}{|\Rset|} \left( \max_{i\in \{1,...,k\} } {\sum_{j\in \Rset \setminus \{i\} }\E[|h_{ij}|^2] } \right)
\end{equation}

\end{theorem}

Before proving Theorem \ref{thm1}, we state Lemma \ref{lemmaconverse} which provides a converse relating the minimum energy per bit to the channel capacity (see also \cite{UnitCost,Zahedi}).

Dropping the time indices, the channel equation \eqref{channeleq} for the received symbol $y_j$ at node $j$ can be rewritten as
\begin{equation}
y_j=\mathbf{h}_j^T\mathbf{x}+z_j
\end{equation}
where $\mathbf{x}=(x_1,...,x_k)^T$ is the transmission symbol vector and $\mathbf{h}_j=(h_{1j},...,h_{(j-1)j},0,h_{(j+1)j},...,h_{kj})^T$ is the vector representing the fading. The complex Gaussian noise $z_j$ is taken to be distributed according to $\mathcal{CN}(0,N_0)$.

\begin{lemma}
\label{lemmaconverse}
For the destination set $\Rset$, the minimum energy per bit for the network satisfies
\begin{equation}
{E_{b}}_{\min} (\Rset) \geq \inf_{\substack{P_1,P_2,...,P_k\geq 0 \\ : \sum_{i=1}^{k}P_i >0 }} \max_{j\in \Rset} \frac{ \sum_{i=1}^{k} P_i}{ \sup_{\substack{P_\mathbf{x}:\\ \E[|x_{i}|^2]\leq P_{i} \text{  for }i= 1,...,k }} I(\mathbf{x};y_{j}|\mathbf{h}_j) } \label{convstatement}
\end{equation}
\end{lemma}
\begin{IEEEproof}
Appendix \ref{prooflemmaconverse}.
\end{IEEEproof}

A brief rationale for Lemma \ref{lemmaconverse} is as following. Pick a node $j$ belonging to the destination set $\Rset$. For the given power constraints --- $P_1,P_2,...,P_k$ on the transmission power, consider the channel from the set of nodes $\{1,...,j-1,j+1,...,k\}$ to node $j$. By the \emph{max-flow min-cut} bound, the rate of reliable communication to node $j$ by the remaining nodes cannot exceed 
\begin{equation}
C_{j}(P_1,...,P_k) \triangleq \sup_{\substack{P_\mathbf{x}:\\ \E[|x_{i}|^2]\leq P_{i}}} I(\mathbf{x};y_{j}|\mathbf{h}_j)
\end{equation}
bits per channel use. Therefore, the number of channel uses per bit is at least $1/C_j(P_1,P_2,...,P_k)$, which implies that the total energy spent per bit in communicating to node $j$ is at least ${\sum_{i=1}^{k}P_i}/{C_j}$. While this energy is spent communicating with node $j$, the transmission may benefit other nodes as well. In general, all the other nodes may be able to decode the message just by listening to the transmissions intended for node $j$. However, the minimum energy required to communicate to node $j$ does not exceed the minimum energy required to communicate to all the nodes in $\Rset$. Therefore, we can lower bound the total energy spent communicating to all the nodes in $\Rset$ by the energy spent communicating to any one of the nodes in $\Rset$.

\begin{IEEEproof}[Proof of Theorem \ref{thm1}]

We can lower bound the minimum energy per bit by
\begin{align}
{E_{b}}_{\min} (\Rset) & \geq \inf_{\substack{\bm{w}:\\ w_i\geq 0, \\ \sum_{i=1}^k w_i=1}} \inf_{P>0} \max_{j\in \Rset} \frac{P}{\left( \sup_{\substack{P_\mathbf{x}:\\ \E[|x_{i}|^2]\leq w_i P}} I(\mathbf{x};y_{j}|\mathbf{h}_j) \right)} \label{csetbefore} \\
& \geq \inf_{\substack{\bm{w}:\\ w_i\geq 0, \\ \sum_{i=1}^k w_i=1}} \max_{j \in \Rset} \inf_{P>0} \frac{P}{\left( \sup_{\substack{P_\mathbf{x}:\\ \E[|x_{i}|^2]\leq w_i P}} I(\mathbf{x};y_{j}|\mathbf{h}_j) \right)} \label{cset} \\
& \geq \min_{\substack{\bm{w}:\\ w_i\geq 0, \\ \sum_{i=1}^k w_i=1}} \max_{j\in \Rset} \frac{N_0 \log_e2}{\sum_{i=1}^{k} \E[|h_{ij}|^2] w_i} \label{14}
\end{align}
where the explanation of the steps \eqref{csetbefore}--\eqref{14} is the following. The inequality \eqref{csetbefore} follows from Lemma \ref{lemmaconverse} by rewriting it so that $P$ is the total power and ${\bm{w}}$ is the fractional split of power over all the nodes. The bound in \eqref{cset} follows from the fact that min-max is greater than or equal to max-min. 
To justify \eqref{14}, note that the mutual information term in \eqref{cset} corresponds to the capacity of a multiple transmit and single receive antenna system, which has been widely studied for additive Gaussian noise channels. We are interested in the case where channel state information is not available at the transmitters. For a given probability distribution on $\mathbf{x}$ (independent of $\mathbf{h}_j$), we can bound the mutual information in \eqref{cset} as
\begin{align}
I(\mathbf{x};\mathbf{h}_j^T\mathbf{x}+z_j|\mathbf{h}_j) & \leq \E \left[\log_2 \left( 1+\frac{1}{N_0} \E[|\mathbf{h}_j^T\mathbf{x}|^2 |\mathbf{h}_j] \right) \right] \label{mutual2} \\
 & \leq \frac{\log_2e}{N_0} \E[|\mathbf{h}_j^T\mathbf{x}|^2] \label{mutual3} \\
 & \leq \frac{\log_2e}{N_0} \sum_{i=1}^{k} \E[|h_{ij}|^2] w_i P \label{cset3}
\end{align}
Note that, given $\mathbf{h}_j$, a constraint on the output $y_j=\mathbf{h}_j^T\mathbf{x}+z_j$ is that $\E[|y_j|^2]\leq \E[|\mathbf{h}_j^T\mathbf{x}|^2 |\mathbf{h}_j]+N_0$; thus, the mutual information in \eqref{cset} is maximized when $y_j$ is Gaussian distributed with the given power constraint, which leads to \eqref{mutual2}; the bound in \eqref{mutual3} is obtained using the simple fact that $\log_e(1+x) \leq x$ for all $x>0$; we obtain \eqref{cset3} by maximizing the right hand side of \eqref{mutual3} among all $\mathbf{x}$ independent of $\mathbf{h}_j$ such that $\E[|x_{i}|^2]\leq w_i P$, taking into account the fact that the channel coefficients are independent with zero mean. 

Now that we have established \eqref{14} note that
\begin{align}
\max_{\substack{\bm{w}:\\ w_i\geq 0,\\ \sum_{i=1}^k w_i=1}} \min_{j\in \Rset}\sum_{i=1}^{k} \E[|h_{ij}|^2] w_i & \leq \max_{\substack{\bm{w}:\\ w_i\geq 0,\\ \sum_{i=1}^k w_i=1}} \frac{1}{|\Rset|}\sum_{j\in \Rset}\sum_{i=1}^{k} \E[|h_{ij}|^2]  w_i  \label{18point5} \\
 & = \max_{\substack{\bm{w}:\\ w_i\geq 0,\\ \sum_{i=1}^k w_i=1}} \frac{1}{|\Rset|}\sum_{i=1}^{k} w_i  \left( \sum_{j\in \Rset \setminus \{i\}} \E[|h_{ij}|^2] \right) \label{14.5} \\
 & = \frac{1}{|\Rset|} \max_{i \in \{1,...,k\}} \left( \sum_{j\in \Rset \setminus \{i\}} \E[|h_{ij}|^2] \right) \label{15} \\
 & = G(\Rset) \label{16}
\end{align}
where \eqref{18point5} is obtained by upper-bounding the minimum by an average; the maximum in \eqref{14.5} is attained when all the weight is put on that $i$ for which $\sum_j\E[|h_{ij}|^2]$ is largest. Substituting the effective network radius term from \eqref{16} in \eqref{14} provides the requisite lower bound on the minimum energy per bit.
\end{IEEEproof}

\emph{Remark 1:} While we expect the minimum energy per bit to increase (hence, the effective network radius to decrease) as the destination set becomes larger, it can be shown that the effective network radius does not always decrease with the size of destination set. Therefore, it is useful to maximize the right hand side of \eqref{thm1statement1} by considering all non-empty subsets of the destination set $\Rset$. Thus, a tighter bound on the minimum energy per bit is given by
\begin{equation}
\EbNomin \Bigl( \Rset \Bigr) \geq \max_{\substack{\Rset' \subset \Rset, \\ \Rset' \neq \phi}} \frac{\log_e2}{G (\Rset')} \label{thm1statement2}
\end{equation}

\emph{Remark 2:} For the simple case of a point-to-point Gaussian channel, the effective network radius is simply the channel gain from the source node to the destination node. The bound is tight in this case \cite[Theorem 1]{Spectral}.

\emph{Remark 3:} While Theorem \ref{thm1} holds for all destination sets, for the networks considered later in this paper, we are interested in the particular multicast setting where destination set is set of all the non-source nodes in the network (also known as \emph{broadcasting}).

\section{Flooding Algorithm}
\label{algoflood}
To derive upper bounds on the minimum energy per bit we use a version of the well known {flooding algorithm}. This algorithm, with suitable parameter values, is used to achieve energy-efficient multicasting for the various networks considered later in the paper. 

Since minimum energy per bit requires very small spectral efficiency even in the point-to-point case, we do not place any bandwidth constraints. Therefore, we can assign each transmitter its own wide frequency band. In the wideband regime, the knowledge of the channel states at the receiver does not decrease the minimum energy per bit \cite{Spectral}. Furthermore, a necessary condition for reliable decoding is that the received energy per bit be greater than $N_0\log_e2$. Various wideband communication schemes can be constructed which let the receivers reliably decode a message if the total received energy per bit exceeds $N_0\log_e2$ \cite{Spectral}.

\subsubsection{Description of the algorithm}
The flooding algorithm consists of two parts: an outer algorithm and an inner coding scheme. The outer algorithm \Flood is the description at the \emph{time slot} level using the \emph{decoding} and \emph{encoding} functionalities provided by the inner scheme. (See Fig. \ref{FloodAlgo}).

Time is divided into slots: $1,2,...,T$, each time slot consisting of enough time to let a node transmit one codeword. Multiple nodes can simultaneously transmit in a slot, albeit in their own mutually orthogonal frequency bands. The transmission process is initiated by the source node which is the only node transmitting in the first slot. In any slot thereafter, whether a non-source node transmits and what it transmits is dependent on when and what the node has decoded so far. In particular, if a node decodes a message for the first time in slot $t$, it retransmits the codeword corresponding to the decoded message in slot $t+1$. The decoding process and the determination of the codeword to be transmitted is handled by the inner coding scheme.

Note that every node transmits either never or once. The total number of slots $T$ in the algorithm is a design parameter that depends on the size of the network. 

\begin{figure}[t]
\centering
\framebox{
\begin{minipage}{0.8\textwidth}
{

{\underline{\Flood }}

\begin{enumerate}
\item The source node transmits only in the $1^{st}$ time slot with energy per bit ${E_b}_1$.

\item At the beginning of time slot $t=2,...,T$, each node (except the source node) executes the following
\begin{itemize}
\item If the node was able to decode a message for the first time in the previous time slot, then it retransmits the same message in the current time slot with energy per bit ${E_b}_2$.
\item Else, keep quiet.
\end{itemize}
\end{enumerate}
}
\end{minipage} }
\caption{The Flooding Algorithm: \Flood}
\label{FloodAlgo}
\end{figure}

\subsubsection{Energy consumption of \Flood}

In a network with $k$ nodes, the source node transmits energy per bit ${E_b}_1$ and each of the rest of the $k-1$ non-source nodes transmit either $0$ or ${E_b}_2$. Therefore, the total energy consumed per information bit by \Flood is at most 
\begin{align}
{E_b}_{\text{total}} \leq {E_b}_1+(k-1){E_b}_2 \label{EBT}
\end{align}

Instead of a single flooding scheme, we will demonstrate a sequence of flooding schemes which achieve a vanishing probability of error. ${E_b}_{\text{flood}}$ will be used to denote the infimum of ${E_b}_{\text{total}}$ over this sequence of schemes. Clearly, ${E_b}_{\text{flood}}$ is an achievable energy per bit for the network and thus an upper bound on the ${E_b}_{\min}$ of the network.

\subsubsection{Inner coding scheme}

The \emph{transmit} operation in \Flood uses identical codebooks for all nodes. The task for each decoder is to observe transmissions over multiple time slots and frequency bands. Using these observations, it forms a reliable estimate of the source message. At the end of each time slot it determines whether it has enough information to decode the message. If not, it keeps quiet and waits for more transmissions. If it is able to decode a message, it re-encodes the decoded message and transmits it in the next slot for the benefit of its peers, and remains quiet after that.

\section{Large Random Networks}

This section is devoted to the analysis of random networks, where the number of nodes $k$ goes to infinity and their locations are random. We focus on the cases of dense and extended networks (see Section \ref{summary}). We obtain bounds on ${E_b}_{\min}$ that hold almost surely in the network topology as $k\rightarrow \infty$.

In both cases, the $k$ nodes are placed over a square of area $\Ak$. The diagonal coordinates of the square are $(0,0)$ and $(\sqrt{\Ak},\sqrt{\Ak})$, and the source node is placed on the coordinate $(0,0)$. This is a least favorable location for the source node but turns out to be irrelevant for the scaling laws we derive.

\subsection{Path Loss Model}
\label{pathlossmodel}

The channel gain $\E[|h_{ij}|^2]$ of the link between nodes $i$ and $j$ is determined by their separation $r_{ij}$. This relation is given by a monotonically decreasing \emph{power gain} or \emph{path loss} function $g(r):\R_+\mapsto \R_+$, i.e.,
\begin{equation}
\E[|h_{ij}|^2]=g(r_{ij})
\end{equation}
where, for all $r\geq r_0$,
\begin{equation}
g(r)=r^{-\alpha} \label{farfieldcase}
\end{equation}
where $r_0>0$ and $\alpha>2$ are constants of the model.

To deal with the near-field case, we also put an upper bound on the gain function --- i.e., there is a constant $\bar{g}>0$ such that 
\begin{equation}
g(0)\leq \bar{g} \label{nearfieldcase}
\end{equation}
since the gain cannot be arbitrarily large. Thus, the path loss model is completely characterized by $\alpha$, $r_0$ and $\bar{g}$.

\subsection{Dense Random Networks}
\label{denserandomnetwork}

\begin{figure}[t]
\centering
\scalebox{0.5}
 {\input{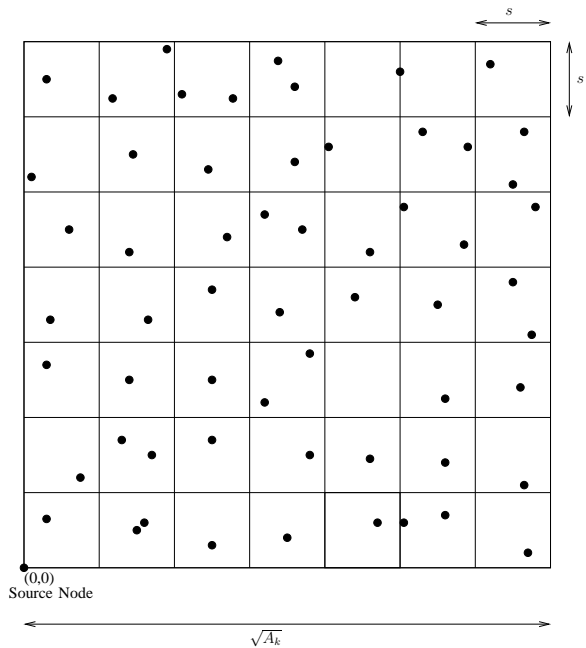}}
\caption{Dense Random Network}
\label{figureDRN}
\end{figure}

A dense random network with $k\geq 2$ nodes consists of a source node at the origin and $k-1$ non-source nodes distributed independently and uniformly over a square of area 
\begin{equation}
\Ak=o\left(\frac{k}{\log k}\right) \label{smallocondition}
\end{equation} 
In addition, we also assume that 
\begin{equation}
{r_0^2}\leq 8\Ak  \label{27p5}
\end{equation}
for all $k\geq 2$.

The results for this case are presented in Theorem \ref{DRNC}, which states that the minimum energy per bit of a dense random network scales linearly with area, almost surely as $k\rightarrow \infty$. The almost sure statement is made with respect to the location of the nodes.

\begin{theorem}
\label{DRNC}
With probability 1, the node placement is such that
\begin{equation}
c_1 \leq \frac{1}{\Ak} {\frac{E_b}{N_0}}_{\min}
\end{equation}
and,
\begin{equation}
\frac{1}{\Ak} {\frac{E_b}{N_0}}_{\text{flood}} \leq c_2
\end{equation}
for all but a finite number of $k$, where
\begin{equation}
c_1 = \frac{2\log_e2}{ 49\, \bar{g}r_0^2+ \frac{2^{\alpha +2}}{\alpha-2} \frac{3}{r_0^{\alpha-2}} } \label{constantc1}
\end{equation} 
and 
\begin{equation}
c_2 = 24\, r_0^{\alpha-2} \log_e2  \label{constantc2}
\end{equation}
\end{theorem}

\begin{IEEEproof}

We begin by partitioning the area $\Ak$ into square cells with side length
\begin{equation}
0<s\leq \frac{r_0}{\sqrt{8}}
\end{equation}
independent of $k$. (See Fig. \ref{figureDRN}). Some of the cells may not be \emph{whole}, i.e. they may not cover an area of $s^2$. However, all these cells would only lie along the upper and the right sides of the square $\Ak$. For simplicity, we restrict our attention to whole cells (i.e., ${\sqrt{\Ak}}$ is a multiple of $s$) for the rest of the proof. Note that there are a total of $\Ak/s^2$ cells in the network. We use ``cell'' to not only refer to the geographical cell but also to the set of nodes falling within the cell. Let $\mathcal{C}$ be the set of cells, and let $\k(\cell)$ denote the number of nodes in cell $\cell \in \mathcal{C}$.

For any $\delta >0$, define a \emph{good placement} event $\mathcal{D}_k$ as the collection of node placement realizations for which all the cells contain at least $(1-\delta)(k-1)s^2/\Ak$ nodes and less than $((1+\delta){(k-1)s^2}/{\Ak})+1$ nodes. Instrumental to both direct and converse parts of the proof, Lemma \ref{gooddist} lower bounds the probability of good placement for a given $k$.

\begin{lemma}
\label{gooddist}
\begin{equation}
\P[{\mathcal{D}}^c_k] \leq \frac{2\Ak}{s^2}\exp\left(-\delta^2(1-\delta)\frac{(k-1)s^2}{2\Ak} \right)
\end{equation}
for all $k\geq 2$.
\end{lemma}
\begin{IEEEproof}
Appendix \ref{proofgooddist}.
\end{IEEEproof}

{\emph{Proof of converse}}

For a given $\delta >0$, let us assume, for the time being, that the event $\mathcal{D}_k$ happens. Also, let 
\begin{equation}
2s \leq {r_0} \label{equationcondition}
\end{equation} 

In order to be able to apply Theorem \ref{thm1}, we need to determine the effective radius $G(\Rset)$ of this network for $\Rset=\{2,...,k\}$. To do so, we first upper bound the quantity $\sum_{j\in \Rset \setminus \{i\} }\E[|h_{ij}|^2]=\sum_{j\in \Rset \setminus \{i\} }g(r_{ij})$ for any node $i\in \{1,...,k \}$. Instead of directly evaluating the total channel gain from node $i$ to all the non-source nodes, we bound it from above by summing the maximum possible gains to all the nodes falling in a cell $\ell$ steps away, for $\ell=0,1,...,({\sqrt{\Ak}}/{s})-1$. Thereafter, it is just a matter of simpifying the terms. Care needs to be exercised, treating those cells falling in the near-field separately from those in the far-field. Therefore, for any node $i \in \{1,...,k\}$,
\begin{align}
\sum_{\substack{j=2\\ j\neq i}}^{k}  g(r_{ij}) & \leq \left( (1+\delta)\frac{(k-1)s^2}{\Ak}+1\right) \left[ \bar{g}+\sum_{\ell =1}^{({\sqrt{\Ak}}/{s})-1} 8\ell \, g((\ell-1)s) \right] \label{cov7} \\
 & \leq \left( (1+\delta)\frac{(k-1)s^2}{\Ak}+1\right) \left[ \left( 1+8 \sum_{\ell=1}^{\lceil {r_0}/{s} \rceil} \ell \right) \bar{g} + \frac{8}{s^\alpha} \sum_{\ell=\lceil {r_0}/{s} \rceil+1}^{({\sqrt{\Ak}}/{s})-1} \frac{\ell}{(\ell-1)^\alpha}  \right] \label{cov2} \\
& \leq  (1+2\delta)\frac{(k-1)s^2}{\Ak}\left[ \left(1+12\left(\frac{r_0}{s}\right)^2\right)\bar{g} + \frac{ 2^{\alpha}}{\alpha-2} \frac{3}{s^2r_0^{\alpha -2}} \right] \label{cov23} \\
& =  (1+2\delta)\frac{(k-1)}{\Ak}\left[ (12.25)\, \bar{g}r_0^2+ \frac{ 2^{\alpha}}{\alpha-2} \frac{3}{r_0^{\alpha-2}} \right] \label{42p5} \\
& \leq  \frac{(k-1)\log_e2}{c_1\Ak} \label{conv}
\end{align}
where
\begin{equation}
c_1=\frac{4\log_e2}{(1+2\delta)\left( 49 \, \bar{g}r_0^2+ \frac{ 2^{\alpha+2}}{\alpha-2} \frac{3}{r_0^{\alpha-2}} \right) } \label{boundonc1}
\end{equation} 
The explanation of the steps \eqref{cov7}--\eqref{conv} is the following. Suppose that node $i$ falls in cell $\cell$. Consider the set of cells exactly $\ell$ horizontal, vertical or diagonal steps away from $\cell$ (the only cell $\ell=0$ steps away is $\cell$ itself). There are at most $\max\{1,8\ell\}$ cells $\ell$ steps away from $\cell$. The channel gain from node $i$ to a node in any cell exactly $\ell$ steps away is not more than $g((\ell-1)s)$ for $\ell\geq 1$. For $\ell=0$, the channel gain is not more than $\bar{g}$. Furthermore, since we assume that the event $\mathcal{D}_k$ occurs, the maximum number of nodes in a cell is less than $((1+\delta){(k-1)s^2}/{\Ak})+1$. This gives us \eqref{cov7}. 
For $\ell\geq \lceil {r_0}/{s} \rceil+1$, since the minimum distance between the nodes is greater than $r_0$, the upper bound on gain is
\begin{equation}
g((\ell -1)s) \leq \frac{1}{s^{\alpha}(\ell-1)^{\alpha}}
\end{equation}
This immediately leads to \eqref{cov2}. Step \eqref{cov23} requires a number of minor simplifications. First, 
\begin{equation}
(1+\delta)\frac{(k-1)s^2}{\Ak}+1 \leq  (1+2\delta)\frac{(k-1)s^2}{\Ak} \label{simpli1}
\end{equation}
for all $k$ large enough. Due to \eqref{equationcondition}, the following inequalities hold.
\begin{equation}
\Bigl\lceil \frac{r_0}{s}\Bigr\rceil+1 < 2 \frac{r_0}{s} \label{valid1}
\end{equation}
\begin{equation}
\Bigl\lceil \frac{r_0}{s} \Bigr\rceil-1 \geq\frac{1}{2} \frac{r_0}{s} \label{valid2}
\end{equation}
and,
\begin{equation}
\Bigl\lceil \frac{r_0}{s} \Bigr\rceil < \frac{3}{2}\frac{r_0}{s}
\end{equation}
Thus, we have the following bound on the first sum in \eqref{cov2}
\begin{equation}
\sum_{\ell=1}^{\lceil {r_0}/{s} \rceil}\ell < \frac{3}{2} \left( \frac{r_0}{s} \right)^2 \label{simpli2}
\end{equation}
since the sum of first $n$ natural numbers is $\frac{1}{2}n(n+1)$. The other sum in \eqref{cov2} can be bounded as
\begin{align}
\sum_{\ell=\lceil {r_0}/{s} \rceil+1}^{({\sqrt{\Ak}}/{s})-1} \frac{\ell}{(\ell-1)^\alpha} & = \sum_{\ell=\lceil {r_0}/{s} \rceil}^{({\sqrt{\Ak}}/{s})-2} \frac{\ell +1}{\ell^\alpha} \label{leq0} \\
& \leq \frac{3}{2} \sum_{\ell=\lceil {r_0}/{s} \rceil}^{\infty} \frac{1}{\ell^{\alpha-1}} \label{leq1} \\
& \leq \frac{3}{2}\int_{\lceil r_0/s\rceil -1}^{\infty}\frac{1}{u^{\alpha-1}}\,du \\
& = \frac{3}{2(\alpha-2)} \frac{1}{\left(\lceil \frac{r_0}{s} \rceil -1 \right)^{\alpha-2}} \label{simpli3}
\end{align}
where we have used the fact that $x +1 \leq \frac{3}{2} x $ for all $x\geq 2$ in \eqref{leq1}. Note that \eqref{simpli3} is valid only for $\alpha>2$. 

Having bounded the total gain from any node $i$ in \eqref{conv}, we get the following upper bound on the effective network radius.
\begin{equation}
G(\Rset) \leq \frac{\log_e2}{c_1\Ak}
\end{equation}
which implies, by Theorem \ref{thm1}, 
\begin{equation}
\EbNomin \geq c_1 \Ak \label{bound231}
\end{equation}
Since the choice of $\delta>0$ in \eqref{boundonc1} is arbitrary, we pick $\delta=1/2$ to get the requisite lower bound on the minimum energy per bit for $c_1$ given by \eqref{constantc1}.

We now show that $\mathcal{D}_k$ occurs almost surely as $k\rightarrow \infty$, which implies the converse part of Theorem \ref{DRNC} since, conditioned on $\mathcal{D}_k$, the bound in \eqref{bound231} holds for all $k$ large enough. Note that the condition \eqref{smallocondition} implies that $\Ak \leq k$ for all $k$ large enough, and that for any constant $c'\geq 0$, the value of $(k-1)/\Ak$ is greater than $c'\log_e k$ for all $k$ large enough. Pick a $k'$ large enough that these three conditions are satisfied (with $c'=6/(\delta^2(1-\delta)s^2)$) for any $k \geq k'$. 
Now, consider the sum
\begin{align}
\sum_{k=2}^{\infty} \P[\mathcal{D}_k^c] & \leq \frac{2}{s^2} \sum_{k=2}^{\infty} \Ak \exp \left(-\delta^2(1-\delta)\frac{(k-1)s^2}{2\Ak} \right) \label{1step1111} \\
 & \leq c+\frac{2}{s^2} \sum_{k=k'}^{\infty} k \exp \left(-\delta^2(1-\delta)\frac{(k-1)s^2}{2\Ak} \right) \label{eq123} \\
 & \leq c+\frac{2}{s^2} \sum_{k=k'}^{\infty} k \exp \left(-3 \log_ek \right) \label{eq124} \\
 & = c+\frac{2}{s^2} \sum_{k=k'}^{\infty} \frac{1}{k^2} \\
 & \leq c+\frac{2}{s^2} \frac{\pi^2}{6} \\
 & < \infty
\end{align}
where $c$ is some real positive constant. Inequality \eqref{1step1111} is due to Lemma \ref{gooddist}; and the inequalities \eqref{eq123}, \eqref{eq124} are due to the choice of $k'$. 

Therefore, since $\sum_{k=2}^{\infty} \P[\mathcal{D}_k^c]$ is finite, by the Borel-Cantelli lemma, we conclude that with probability 1 the event $\mathcal{D}_k^c$ occurs only a finite number of times. Hence, $\P[\lim_{k \rightarrow \infty} \inf \mathcal{D}_k]=1$.

{\emph{Proof of achievability}}

For any $\epsilon_1,\epsilon_2>0$, we first show that 
\begin{equation}
\text{\texttt{FLOOD}}\left(\frac{N_0\log_e2}{g(\sqrt{8}s)}+\epsilon_1,\frac{(1+\epsilon_2) \Ak N_0\log_e2}{s^2(k-1)g(\sqrt{8}s)} \right) \label{flooddense}
\end{equation}
manages to reliably communicate the common message to all the nodes, conditioned on the event $\mathcal{D}_k$.

Set $s=r_0/\sqrt{8}$ so that we can replace $g(\sqrt{8}s)$ with $r_0^{-\alpha}$. Thus, the total energy consumption per bit of \eqref{flooddense} is  
\begin{align}
{E_b}_{\text{total}} 
 & \leq \left( r_0^{\alpha}+ (1+\epsilon_2) 8 \Ak r_0^{\alpha-2} \right)N_0 \log_e2 + \epsilon_1 \label{63p00}\\
 & \leq \left((2+\epsilon_2)8r_0^{\alpha-2}\log_e2  \right)\Ak N_0 + \epsilon_1 \label{65p00}
\end{align}
where we have used the lower bound \eqref{27p5} on $\Ak$ in simplifying \eqref{63p00} to \eqref{65p00}. 

Our next step is to show that the scheme in \eqref{flooddense} is able to reach all the nodes.
First, for the given value of $\epsilon_2$, we choose any 
\begin{equation}
0 < \delta < \frac{\epsilon_2}{2(1+\epsilon_2)} \nonumber
\end{equation} 
in the definition of event $\mathcal{D}_k$. Thus, if $\mathcal{D}_k$ occurs, then all the cells have at least 
\begin{equation}
\frac{1+\frac{\epsilon_2}{2} }{1+\epsilon_2}\frac{(k-1)r_0^2}{8\Ak} \label{minnodes}
\end{equation} 
nodes.

Let $T_k={\sqrt{8\Ak}}/{r_0}$ be the maximum number of time slots in the flooding scheme. Suppose that a non-source node $i$ belongs to cell $\cell$. For any cell $\cell$, there is a sequence $( \cell_1,\cell_2,...,\cell_T )$ of horizontally, vertically or diagonally adjacent cells such that $\cell_1$ is the origin cell \footnote{\emph{Origin cell} is the cell containing the source node.} and $\cell_T=\cell$, for some $T \leq T_k$. We now present an argument to show that the nodes in cell $\cell_t$ successfully decode by the end of slot $t-1$.

In the first time slot, the source node transmits with energy per bit 
\begin{equation}
E_{b_1} > {r_0^{\alpha}N_0\log_e2}
\end{equation}
which implies that the received energy per bit at all the nodes within a radius of $r_0$ (which includes all the nodes falling within two cells of the origin) is strictly greater than $N_0\log_e2$. Therefore, all the nodes in the cell $\cell_2$ are able to decode the message reliably with vanishing probability of error \cite[Theorem 1]{Spectral}. Now, suppose that all the nodes in cell $\cell_{t+1}$ are able to decode without error by the end of slot $t$. This implies that all the nodes in the cell $\cell_{t+1}$ would have transmitted the correct message by the end of slot $t+1$ (possibly in different slots). The energy per bit of their transmissions is 
\begin{equation}
E_{b_2} = \frac{(1+\epsilon_2) 8 \Ak r_0^{\alpha -2} N_0 \log_e2}{k-1}
\end{equation}
For any node $i$ in the cell $\cell_{t+2}$, the gain from any of the nodes in $\cell_{t+1}$ is at least $r_0^{-\alpha}$ since the distance between them is at most $r_0$. Furthermore, the minimum number of nodes in the cell $\cell_{t+1}$ is given by \eqref{minnodes}. Therefore, the total received energy at node $i$ is 
\begin{align}
E_b^r & \geq r_0^{-\alpha} \left(\frac{1+\frac{\epsilon_2}{2}}{1+\epsilon_2} \frac{(k-1)r_0^2}{8\Ak}\right) \left(\frac{(1+\epsilon_2) 8 \Ak r_0^{\alpha-2} N_0 \log_e2}{k-1}\right) \\
 & > N_0 \log_e2
\end{align}
Moreover, all this energy is received by the end of slot $t+1$. Therefore, the node $i$ (and hence, every node in $\cell_{t+2}$) successfully decodes the message by the end of slot $t+1$ with vanishing probability of error. Thus, by inductive reasoning, the flooding scheme is able to reach out to every cell $\cell$ (and hence, to every node) by the end of slot $T_k-1$.

Note that since the choice of $\epsilon_1$ and $\epsilon_2$ is arbitrary, from \eqref{65p00} we get that ${E_b}_{\text{flood}}$ satisfies the condition laid out in the statement of Theorem \ref{DRNC} with the constant $c_2$ given by \eqref{constantc2}, for $\epsilon_1 \rightarrow 0$ and $\epsilon_2=1$.

In our analysis above, we depend on the occurrence of $\mathcal{D}_k$ in order to ensure that each cell contains at least the number of nodes given in \eqref{minnodes}. By the same argument as in the converse, we get that the direct bound holds for all but a finite number of $k$.
\end{IEEEproof}

\emph{Remark:} If $\alpha=2$, the sum on the left hand side of \eqref{leq0} would grow as $\Theta(\log \Ak)$ which would weaken (\ref{conv}) to a constant times ${(k-1)\log \Ak}/{\Ak}$. Therefore, the bound now becomes 
\begin{equation}
\EbNomin \geq c'_1 \frac{\Ak}{\log \Ak}
\end{equation}
for some constant $c'_1>0$. This was shown to be achievable in \cite{ScaglioneJSAC} in a different setting (as mentioned in Section \ref{priorwork}).

\subsection{Random Extended Networks}
\label{randomextendednetwork}
The extended random network case differs from the dense case because the density of the nodes is now a constant:
\begin{equation}
\lambda =\frac{k}{\Ak} \label{lambdarelation}
\end{equation}
in nodes/$\text{m}^2$.

As in dense networks, the source node is placed at the origin and the rest $k-1$ non-source nodes are distributed independently and uniformly over the area $\Ak$.

\begin{figure}
\centering
\scalebox{0.5}
 {\input{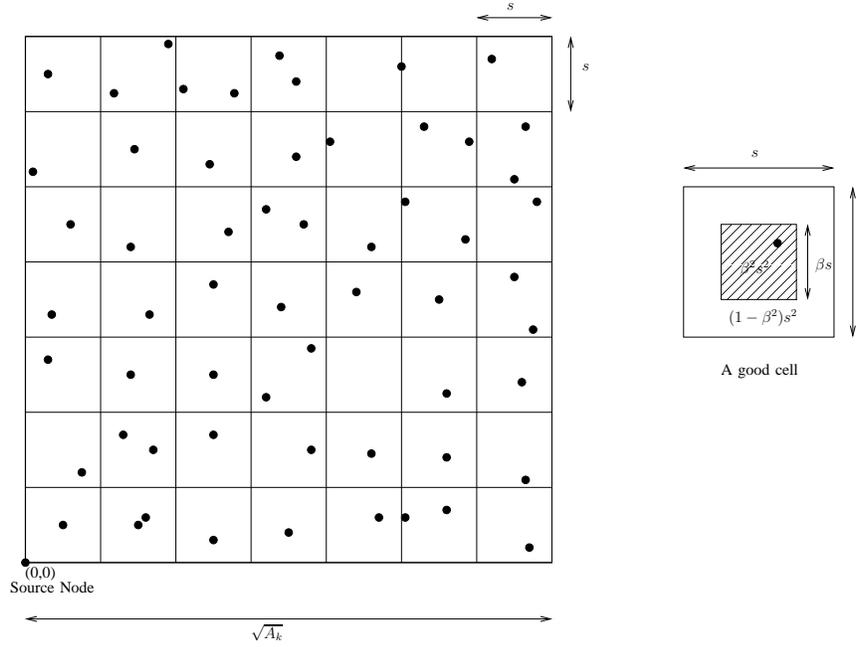}}
\caption{Random extended network}
\label{figure1}
\end{figure}

The result for this case is presented in Theorem \ref{thm-extended}.

\begin{theorem}
\label{thm-extended}
With probability 1, the node placement is such that
\begin{equation}
c_1 \leq \frac{1}{k} {\frac{E_b}{N_0}}_{\min} \label{thm-extendedbound1}
\end{equation}
and,
\begin{equation}
\frac{1}{k(\log_e k)^{\alpha/2}} {\frac{E_b}{N_0}}_{\text{flood}} \leq c_2 \label{thm-extendedbound2}
\end{equation}
for all but a finite number of $k$, where
\begin{equation}
c_1 = \left\{ \begin{array}{ll} 
\frac{\log_e2}{2^43^{\alpha+2}e\zeta(\alpha-1)}\lambda^{-\alpha/2} & \text{for } \lambda < \frac{1}{9r_0^2} \\
\frac{\log_e2}{2^53^3\left(\bar{g}r_0^2+ \frac{1}{(\alpha-2)6^{\alpha}r_0^{\alpha-2}} \right)}\lambda^{-1} & \text{for } \lambda \geq \frac{1}{9r_0^2}
\end{array} \right. \label{constantc3}
\end{equation}
where $\zeta(\cdot)$ is the Riemann zeta function.
And,
\begin{equation}
c_2 = 3\, 2^{2\alpha-1}(\log_e2) \lambda^{-\alpha/2} \label{constantc4}
\end{equation}

\end{theorem}

\begin{IEEEproof}

\emph{Proof of converse}

For simplicity, we assume that the number of nodes $k$ is a square integer larger than 1.
Partition the network area into square cells with side length $\lambda^{-1/2}$. This implies that there are $k$ cells in the network. Let $\mathcal{C}$ be the set of cells, and let $\k(\cell)$ denote the number of nodes in cell $\cell \in \mathcal{C}$. Next, right at the center of each cell, consider a small square \emph{window} of side length $\beta \lambda^{-1/2}$, where $0\leq \beta \leq 1$ is a constant to be selected later. Define a non-origin cell to be \emph{good} if it contains exactly one node within its window and no nodes outside the window. The non-source node falling in a good cell is called a \emph{good node}. (See Fig. \ref{figure1}). Let the set of good nodes be $\Rset_1\subset \{2,...,k\}$. The number of good nodes (cells) is denoted by $k_1$. The set $\Rset_1$ is our destination set, for which we will calculate the effective network radius $G(\Rset_1)$.

For any $\delta > 0 $, define a \emph{good placement} event $\mathcal{D}_k$ as the collection of node placements for which
\begin{equation}
 k_1 \geq  (1-\delta) \beta^2 \frac{(k-1)^k}{k^{k-1}} \label{boundk1}
\end{equation}
The next result lower bounds the probability of a good placement.

\begin{lemma}
\label{lemma3p0}
\begin{equation}
\label{appmcdiarmid}
\P[\mathcal{D}_k^c] \leq 2\exp\left(- \frac{1}{2}\delta^2\beta^4 \frac{(k-1)^{2k-1}}{k^{2k-2}} \right)
\end{equation}
for all $k\geq 2$.
\end{lemma}
\begin{IEEEproof} 
Appendix \ref{prooflemma3p0}.
\end{IEEEproof}

Let us assume, for the time being, that the event $\mathcal{D}_k$ happens and focus our attention on finding an upper bound on $G(\Rset_1)$. For any node $i \in \{1,...,k\}$, we have the following upper bound on total gain from node $i$ to the nodes in $\Rset_1$.
\begin{align}
\sum_{j \in \Rset_1\setminus \{i\}} g(r_{ij}) & \leq \sum_{\ell =1}^{\sqrt{k} -1} 8\ell\, g(r_{ij}) \label{1step1} \\
 & \leq \sum_{\ell =1}^{\sqrt{k} -1} 8\ell\, g\left( \frac{(\ell -1)}{\sqrt{\lambda}}+\frac{1-\beta}{2\sqrt{\lambda}} \right) \label{1step2}
\end{align}
where \eqref{1step1} and \eqref{1step2} are obtained as follows. Consider all those good cells lying exactly $\ell$ steps (horizontally, vertically or diagonally) away from the cell containing node $i$. Since there are at most $8\ell$ cells $\ell$ steps away, there are at most as many good cells $\ell$ steps away, implying \eqref{1step1}. Furthermore, a good node $\ell$ steps away is at a distance of $((\ell-1)/\sqrt{\lambda})+({(1-\beta)}/{2\sqrt{\lambda}})$ or greater. Thus, the gain to that cell cannot exceed $g\left( ((l-1)/\sqrt{\lambda})+({(1-\beta)}/{2\sqrt{\lambda}}) \right)$, implying \eqref{1step2}. 

To further simplify the right hand side of \eqref{1step2}, we need to consider the following two cases for $\lambda$,
\begin{enumerate}
\item $\lambda < {1}/{(9r_0^2)}$

Let $\beta=1/3$. Continuing from \eqref{1step2}, an upper bound on the total gain from node $i$ to the nodes in $\Rset_1$ is
\begin{align}
\sum_{j \in \Rset_1\setminus \{i\}} g(r_{ij}) & \leq \sum_{\ell =1}^{\infty} 8\ell\, g\left(\frac{1-\beta}{2\sqrt{\lambda}}\ell \right) \label{2step2} \\
 & \leq 8\left(\frac{1-\beta}{2\sqrt{\lambda}} \right)^{-\alpha} \sum_{\ell =1}^{\infty} \ell^{-(\alpha-1)} \label{2step3}\\
 & = 2^3\, 3^{\alpha}\zeta (\alpha-1) \lambda^{\alpha/2} \label{2step4}
\end{align}
where, in \eqref{2step4}, $\zeta(\cdot)$ is the Riemann zeta function which is finite for all real arguments greater than 1; \eqref{2step2} is obtained from \eqref{1step2} by observing that $((\ell-1)/\sqrt{\lambda})+({(1-\beta)}/{2\sqrt{\lambda}})\geq {(1-\beta)}\ell /{2\sqrt{\lambda}}$ for all $\ell \geq 1$. Also, ${(1-\beta)}/{2\sqrt{\lambda}}> r_0$ for the given value of $\beta$, due to the defining condition of this case. Therefore, all the gain terms are given by the far-field case \eqref{farfieldcase}, which immediately implies \eqref{2step3}. 

Since the bound in \eqref{2step4} holds for every node $i\in \{1,...,k\}$, from \eqref{boundk1} and \eqref{2step4} we get the following upper bound on effective network radius
\begin{align}
G(\Rset_1) & \leq \frac{2^3\, 3^{\alpha}\zeta (\alpha-1) \lambda^{\alpha/2}}{(1-\delta) \beta^2 \frac{(k-1)^k}{k^{k-1}}} \\
 & \leq \frac{2^3\, 3^{\alpha+2}e\zeta (\alpha-1) \lambda^{\alpha/2}}{(1-2\delta) k} \label{GR1}
\end{align}
for all $k$ large enough. Inequality \eqref{GR1} is due to the fact that since $\lim_{k \rightarrow \infty} \left(1-1/k \right)^k=1/e$, for any $\delta>0$
\begin{equation}
\left(1-\frac{1}{k} \right)^k \geq \left( \frac{1-2\delta}{1-\delta}\right)\frac{1}{e}
\end{equation}
for all $k$ large enough.

Using \eqref{GR1} in Theorem \ref{thm1} immediately gives us that for all $k$ large enough
\begin{equation}
\EbNomin \geq (1-2\delta) \left( \frac{\log_e2} {2^{3}3^{\alpha+2}e\zeta(\alpha-1)}  \right) k\lambda^{-\alpha/2}
\end{equation}
Since the choice of $\delta>0$ is arbitrary, taking $\delta = 1/4$ we obtain the claimed bound for $\lambda < 1/(9r_0^2)$.

\item $\lambda\geq {1}/{(9r_0^2)}$

Now, let $\beta=1$. 

Define 
\begin{equation}
L \triangleq \lceil {6r_0\sqrt{\lambda}} \rceil
\end{equation}
which is roughly the number of cells beyond which the far-field model is valid. Note that $L\geq 2$.

Continuing from \eqref{1step2}, an upper bound on the total gain from node $i$ to the nodes in $\Rset_1$ is
\begin{align}
\sum_{j \in \Rset_1\setminus \{i\}} g(r_{ij}) & \leq \sum_{\ell =1}^{L} 8\ell\, \bar{g} + \sum_{\ell=L+1}^{\sqrt{k} -1} 8\ell\, g\left( \frac{(\ell-1)}{\sqrt{\lambda}} \right) \label{3step1}\\
 & \leq 8\bar{g} \sum_{\ell=1}^{L} \ell + 8{\lambda}^{\alpha/2} \sum_{\ell =L+1}^{\infty} \frac{\ell}{(\ell-1)^{\alpha}} \label{3step2}\\
 & \leq 4L(L+1)\bar{g} + 12 {\lambda}^{\alpha/2} \sum_{\ell =L}^{\infty} {\ell}^{-(\alpha-1)} \label{3step3}\\
 & \leq 12 \left({6r_0\sqrt{\lambda}} \right)^2\bar{g} + \frac{12{\lambda}^{\alpha/2}}{\alpha-2} \frac{{\lambda}^{1-(\alpha/2)}}{6^{\alpha-2}r_0^{\alpha-2}} \label{3step4}\\
 & = 2^4\, 3^3 \left( r_0^2\bar{g} + \frac{1}{(\alpha-2)2^{\alpha}3^{\alpha}r_0^{\alpha-2}} \right) \lambda \label{3step5}
\end{align}
The first step is obtained from \eqref{1step2} by breaking the sum into two parts, and bounding the gain terms in the first sum by $\bar{g}$ and in the second sum by $g\left( (\ell-1)/\sqrt{\lambda} \right)$; \eqref{3step2} is due to the fact that for $\ell \geq L+1$, the distance $(\ell-1)/\sqrt{\lambda}$ is greater than $r_0$; the explanation for \eqref{3step3} and \eqref{3step4} is similar to that of \eqref{cov7} -- \eqref{conv}.

Since \eqref{3step5} is valid for every node $i\in \{1,...,k\}$, we can bound $G(\Rset_1)$ and thus find the lower bound for this case in a manner similar to the previous case.

\end{enumerate}

To complete the proof of the converse, we still need to show that the converse bound holds for all but a finite number of $k$, with probability 1. To do so, consider the sum
\begin{align}
\sum_{k=2}^{\infty} P[\mathcal{D}_k^c] & \leq 2 \sum_{k=2}^{\infty} \exp \left( -\frac{1}{2}\delta^2 \beta^4 \frac{(k-1)^{2k-1}}{k^{2k-2}} \right) \label{4step1}\\
 & \leq 2 \sum_{k=2}^{\infty} \left( \exp \left({- \frac{\delta^2}{2\, 3^4} \left( 1-\frac{1}{k} \right)^{2k-2}}\right) \right)^{k-1} \label{4step2}\\
 & \leq 2 \sum_{k=2}^{\infty} c^{k-1} \label{4step3}\\
 & < \infty \label{4step4}
\end{align} 
where \eqref{4step1} is from Lemma \ref{lemma3p0}; inequality \eqref{4step2} is valid for $\beta=1/3$ or $1$ (which covers both cases); the bound in \eqref{4step3} is obtained by noticing that $(1-1/k)^{2(k-1)}$ is greater than $1/e^2$ for all $k\geq 2$. The value of $c$ in \eqref{4step3} is given by
\begin{equation}
c=\exp \left({- \frac{\delta^2}{2\, 3^4\, e^2}}\right)
\end{equation}
Since $c<1$, the infinite geometric progression in \eqref{4step3} converges to a finite value. In view of the summability of $\P[\mathcal{D}_k^c]$, we can invoke the Borel-Cantelli lemma to conclude that with probability 1 the event $\mathcal{D}_k^c$ occurs only finitely many times. Also, since the occurence of $\mathcal{D}_k$ implies that our converse bound holds for all $k$ large enough, we conclude that with probability 1 the converse bound holds for all but a finite number of $k$.

\emph{Proof of achievability}

For any $\epsilon>0$, consider the scheme
\begin{equation}
\text{\texttt{FLOOD}} \left( \frac{N_0\log_e2}{g(\sqrt{8}s_k)} + \frac{\epsilon}{k}, \frac{N_0\log_e2}{g(\sqrt{8}s_k)} + \frac{\epsilon}{k} \right) \label{multihopextended}
\end{equation}
where $s_k>0$ (to be selected later) is the side length of the cells in the network. For the sake of simplicity, $\sqrt{\Ak}$ is assumed to be a multiple of $s_k$.

Denote by $\mathcal{E}_k$ the event that no cell is empty. Let us assume, for the time being, that $\mathcal{E}_k$ occurs. 
Since all the cells are non-empty, for every cell $\cell$ we have a sequence (of length at most ${\sqrt{\Ak}}/{s_k}$) of adjacent (horizontally, vertically or diagonally) non-empty cells which begins at the cell containing the origin and terminates at cell $\cell$. Therefore, there is a path of nodes from the source node to any other node such that two consecutive nodes are within a distance of $\sqrt{8}s_k$ of each other. This implies that if a node that has already decoded a message transmits with energy per bit greater than ${N_0\log_e2}/{g(\sqrt{8}s_k)}$, its transmission will be received by nodes in the neighboring cells with sufficient energy to decode the message.
Thus, the multi-hopping scheme \eqref{multihopextended} suffices to reach every node. 

For any $\delta>0$, set 
\begin{equation}
s_k^2=\frac{(2+\delta) \log_e\Ak}{\lambda} \label{98p5}
\end{equation}
Since $s_k$ grows unbounded with $k$, $\sqrt{8}s_k\geq r_0$ eventually. Therefore, we can replace $g(\sqrt{8}s_k)$ with
\begin{equation}
\frac{1}{{(\sqrt{8}s_k)}^{\alpha}}=\frac{\lambda^{\alpha/2}} {{(8(2+\delta)\log_e \Ak)}^{\alpha/2}}
\end{equation}
for all $k$ large enough. Thus, from \eqref{EBT} the ${E_b}_{\text{total}}$ of this algorithm satisfies
\begin{equation}
{E_b}_{\text{total}} \leq \frac{{(8(2+\delta)\log_e \Ak)}^{\alpha/2}kN_0 \log_e2}{\lambda^{\alpha/2}}+{\epsilon} 
\end{equation}
Substituting the value of $\Ak$ from \eqref{lambdarelation} and noting that the choices of $\epsilon$ and $\delta$ are arbitrary, we immediately get that 
\begin{equation}
{\frac{E_b}{N_0}}_{\text{flood}} \leq c_2\, k {\lambda}^{-\alpha/2} (\log_e k)^{\alpha/2}
\end{equation}
for all $k$ large enough and $c_2$ as given in \eqref{constantc4}.

Hence, for all $k$ large enough, every node placement in $\mathcal{E}_k$ satisfies the bound \eqref{thm-extendedbound2}. We now show that $\mathcal{E}_k$ occurs almost surely as $k\rightarrow \infty$. For every $k$, let $\cell_1$ be any fixed non-source cell. Consider the sum
\begin{align}
\sum_{k=2}^{\infty} \P[\mathcal{D}_k^c] & = \sum_{k=2}^{\infty} \P[\exists \cell \in \mathcal{C}: \k(\cell)=0] \label{102p5} \\
 & \leq \sum_{k=2}^{\infty} \left( \frac{\Ak}{s_k^2}-1 \right) \P[ \k(\cell_1)=0] \label{eq875} \\
 & \leq \sum_{k=2}^{\infty} \frac{\Ak}{s_k^2} \left( 1-\frac{s_k^2}{\Ak} \right)^{k-1} \label{eq876} \\
 & \leq \sum_{k=2}^{\infty} \frac{k}{(2+\delta)\log \Ak }\exp \left({-(2+\delta)\frac{k-1}{k}\log \Ak}\right) \label{94.5} \\
 & \leq c+\sum_{k=k_1}^{\infty} \frac{k}{(2+\delta)\log \Ak }\, \frac{1}{\Ak^{2+\delta/2}} \label{eq877} \\
 & < \infty
\end{align}
where \eqref{eq875} is the union bound over all the non-origin cells in the network; since the probability of the non-origin cell $\cell_1$ being empty is $\left(1-({s_k^2}/{\Ak}) \right)^{k-1}$, we obtain \eqref{eq876}; noting that $1-x\leq \exp (-x)$ for all $x\geq 0$ and together from \eqref{lambdarelation} and \eqref{98p5}, we obtain \eqref{94.5}; since for any given $\delta>0$, 
\begin{equation}
(2+\delta)\, \frac{k-1}{k} \geq 2+\frac{\delta}{2}
\end{equation}
for all $k\geq k_1$ for all large enough $k_1$, \eqref{94.5} simplifies to \eqref{eq877}, where $c<\infty$ is some constant; the series in \eqref{eq877} converges since $\Ak$ is linear in $k$, and $\sum_{k=1}^{\infty}{k^{-(1+\delta_1)}}$ converges for any $\delta_1>0$.
Therefore, by the Borel-Cantelli lemma, event $\mathcal{E}_k$ occurs for all but a finite number of $k$. This immediately implies the achievability part of Theorem \ref{thm-extended}.
\end{IEEEproof}

\section{Regular networks}
\label{regularnetwork}
In both the dense and the extended random networks, we saw that the proposed bounds on ${E_b}_{\min}$ hold almost surely as $k\rightarrow \infty$. Deviation from these bounds is due to non-favorable placement of nodes, the probability of which is non-zero when $k$ is finite. In this section, we consider finite networks where there is some regularity in node placement. As we show, adhering to the path loss model of Section \ref{pathlossmodel}, even simple regularity conditions allow deterministic results for finite networks.

\begin{figure}[t]
\centering
\scalebox{0.5}
 {\input{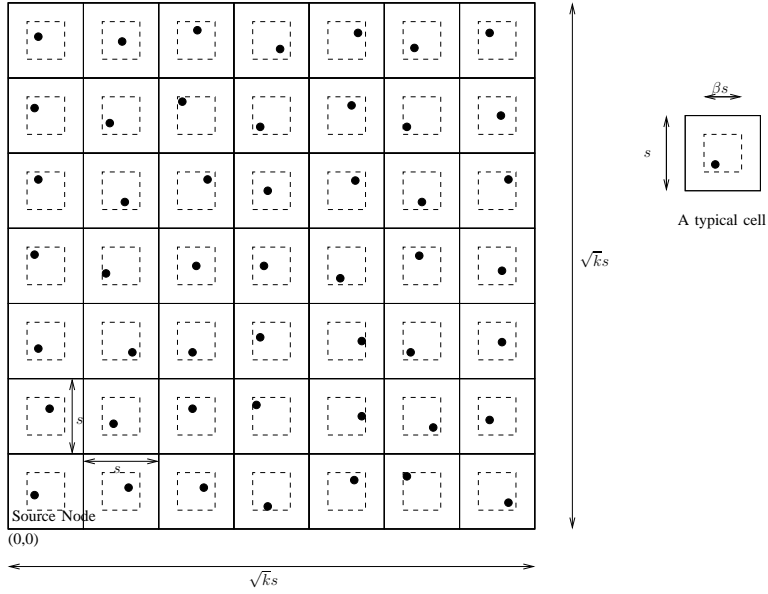}}
\caption{A Regular Network}
\label{figureRN}
\end{figure}

In regular networks, the network area is divided into square cells with side length $s$. 
Each cell has a square window in its center. The window is assumed to occupy a fraction $0\leq \beta^2 < 1$ of the cell area. The regularity condition is that each cell contains exactly one node in its window and no other node outside the window. Each node can be arbitrarily placed within its window. Note that the number of cells is the same as the number of nodes $k$. The source node lies in the window of the origin cell. (See Fig. \ref{figureRN}). For the sake of simplicity, $\sqrt{k}$ is an integer larger than 1. For $x,y=0,1,...,\sqrt{k}-1$, the notation $\cell (x,y)$ is used to denote the cell with the lower left corner on the coordinates $(xs,ys)$. As discussed, the node in cell $\cell (x,y)$ lies within the square having its diagonal coordinates at $(xs+((1-\beta)s/2), ys+((1-\beta)s/2))$ and $(xs+((1+\beta)s/2), ys+((1+\beta)s/2))$.

$\beta$ denotes the flexibility in the placement of the nodes. $\beta=0$ is the case when there is no flexibility and all the nodes fall exactly on the lattice points. 

The following result provides an upper bound on the ratio ${E_b}_{\text{flood}}/{E_b}_{\min}$ that is independent of the number of nodes and the cell size.

\begin{theorem}
\label{RDN}
For any regular network with $k\geq 2$ nodes and cell size $s>0$, 
\begin{equation}
\frac{{E_b}_{\text{flood}}}{{E_b}_{\min}}\leq c_1
\end{equation}
where,
\begin{equation}
c_1 \leq \max \left\{ \frac{2^{\frac{3\alpha}{2}+4}\zeta(\alpha -1)}{(1-\beta)^{\alpha}}, (2r_0)^{\alpha}\bar{g}, \frac{2^{\frac{3\alpha}{2}+3}r_0^{\alpha}}{(1-\beta)^{\alpha+2}}\left( \bar{g}+\frac{2^{\alpha-2}}{(\alpha-2)r_0^{\alpha}} \right)  \left( 1+ 4(1-\beta)^2 \right) \right\} \label{c_1bound}
\end{equation}
\end{theorem}

\begin{IEEEproof}

The analysis is divided into three separate cases.
\begin{enumerate}
\item ${r_0}  < {(1-\beta)s}$.
\item $(k-1)s^2< r_0^2$.
\item $(1-\beta) s \leq r_0 \leq \sqrt{k-1} s$
%\item ${r_0} \geq {(1-\beta)s}$ and $(k-1)s^2 \geq r_0^2$.
\end{enumerate}
In each of the cases above, we derive a lower bound on $\EbNomin$, and also propose a flooding algorithm with suitable parameters to get an upper bound on $\frac{E_b}{N_0}_{\text{flood}}$.

{\emph{Case 1}}

Let us first upper bound $\sum_{j\in \Rset \setminus \{i\}}g(r_{ij})$ for any node $i$ and the destination set $\Rset=\{2,...,k\}$. Begin by noticing that the number of nodes within $\ell$ steps (horizontal, vertical or diagonal) of any node is at most $8\ell$. Moreover, the distance to any node in a cell $\ell$ steps away is at least $(\ell-1)s+(1-\beta)s \geq \ell (1-\beta)s$ and there are at most $\sqrt{k}-1$ steps in any direction. Therefore, for node $i$
\begin{align}
\sum_{j\in \Rset \setminus \{i\}} g(r_{ij}) & \leq \sum_{\ell =1}^{\sqrt{k}-1} 8\ell\, g(\ell (1-\beta)s ) \nonumber \\
 & = 8 \sum_{\ell =1}^{\infty} \frac{\ell}{\ell^{\alpha} (1-\beta)^{\alpha} s^{\alpha} } \nonumber \\
 & \leq \frac{8 \zeta (\alpha-1)}{(1-\beta)^{\alpha} s^{\alpha}} \label{eq1}
\end{align}

Hence, the effective network radius satisfies
\begin{equation}
G (\Rset) \leq \frac{8\zeta (\alpha-1)}{(1-\beta)^{\alpha}s^{\alpha}(k-1) }
\end{equation}

Therefore, 
\begin{equation}
{\frac{E_b}{N_0}}_{\min}\geq \frac{(1-\beta)^{\alpha} s^{\alpha} (k-1)\log_e2}{8 \zeta (\alpha-1)}
\end{equation}
by Theorem \ref{thm1}.

Next, we note that any node has a sequence (of length at most $\sqrt{k}$) of adjacent (horizontal, vertical or diagonal) cells that begins at the origin cell and ends at the cell containing the node. This translates into a sequence of nodes such that any two adjacent nodes are within a distance $\sqrt{8} s$ of each other. Thus, the multihopping scheme 
\begin{equation}
\text{\texttt{FLOOD}} \left(\frac{N_0 \log_e2}{g(\sqrt{8}s)}+\epsilon_1, \frac{N_0 \log_e2}{g(\sqrt{8}s)}+\epsilon_1 \right) \label{multi3}
\end{equation}
for any $\epsilon_1>0$, would work well. The energy consumption of this scheme is at most $\left( {kN_0\log_e2}/{g(\sqrt{8}s)}\right) +k\epsilon_1$. Since $\sqrt{8}s>r_0$ by the defining condition for this case, we have $g(\sqrt{8}s)={8^{-\alpha/2}s^{-\alpha}}$.

Therefore, the energy consumption per bit of \eqref{multi3} is bounded as
\begin{equation}
{E_b}_{\text{total}} \leq \frac{2^{\frac{3\alpha}{2}+4}\zeta(\alpha -1)}{(1-\beta)^{\alpha}}\, {E_b}_{\min}+\epsilon
\end{equation}
for any $\epsilon>0$. Therefore, ${E_b}_{\text{flood}}$ which is the infimum of all the achievable ${E_b}_{\text{total}}$ values satisfies
\begin{equation}
{E_b}_{\text{flood}} \leq \frac{2^{\frac{3\alpha}{2}+4}\zeta(\alpha -1)}{(1-\beta)^{\alpha}}\, {E_b}_{\min}
\end{equation}

{\emph{Case 2}}

If $(k-1)s^2< r_0^2$, then $ks^2 < 2r_0^2$ for any $k>1$. This implies that the network can be contained within a square of side $\sqrt{2}r_0$. Therefore, the maximum distance between any two nodes is at most $2r_0$. So, any node can be reached by the one shot transmission scheme
\begin{equation}
\text{\texttt{FLOOD}} \left(\frac{N_0 \log_e2}{g(2r_0)}+\epsilon_1, 0 \right)
\end{equation}
for any $\epsilon_1>0$. Note that $g(2r_0)=2^{-\alpha}r_0^{-\alpha}$.

For the lower bound on energy per bit, note that the gain to any node cannot exceed $\bar{g}$ and there are $k-1$ destination nodes. Therefore, the effective network radius is less than $\bar{g}$. So, no scheme can do better than 
\begin{equation}
\EbNomin \geq \frac{\log_e2}{\bar{g}}
\end{equation}
by Theorem \ref{thm1}.
This immediately implies that 
\begin{equation}
{E_b}_{\text{total}} \leq 2^{\alpha}r_0^{\alpha}\bar{g} \, {E_b}_{\min}+\epsilon
\end{equation}
for any $\epsilon >0$. Therefore,
\begin{equation}
{E_b}_{\text{flood}} \leq 2^{\alpha}r_0^{\alpha}\bar{g} \, {E_b}_{\min}
\end{equation}

{\emph{Case 3}}

Define 
\begin{equation}
\l \triangleq \Big\lfloor \frac{r_0}{(1-\beta)s} \Big\rfloor
\end{equation} 

Since $\l\geq 1$, we have the following bounds on $\l$ :
\begin{equation}
\frac{1}{2}\frac{r_0}{(1-\beta)s} < \l \leq \frac{r_0}{(1-\beta)s}
\end{equation}
and,
\begin{equation}
\frac{r_0}{(1-\beta)s}< \l +1 \leq \frac{2r_0}{(1-\beta)s}
\end{equation}

Next, by the same argument as in case 1,
\begin{equation}
\sum_{j\in \Rset \setminus \{i\}} g(r_{ij}) \leq \sum_{\ell =1}^{\sqrt{k}-1} 8\ell \, g(\ell (1-\beta)s ) \label{1two3}
\end{equation}
Continuing on \eqref{1two3}, we get the following steps. The explanation of these steps is similar to that of \eqref{cov2}--\eqref{conv}. 
\begin{align}
\sum_{j\in \Rset \setminus \{i\}} g(r_{ij}) & \leq 8 \sum_{\ell =1}^{\l} \ell\, g(\ell (1-\beta) s )+ 8 \sum_{\ell=\l +1}^{\infty} \ell\, g(\ell (1-\beta) s) \nonumber \\
 & \leq 4\bar{g} \l (\l+1) + 8 \sum_{\ell=\l +1}^{\infty}  \frac{\ell}{\ell^{\alpha}(1-\beta)^{\alpha} s^{\alpha} } \nonumber \\%\label{tp1} \\
 & \leq 4\bar{g} \frac{2r_0^2}{(1-\beta)^2s^2}+\frac{8}{(1-\beta)^{\alpha}(\alpha-2)}\frac{1}{s^{\alpha}}\frac{1}{{\l}^{\alpha-2}} \nonumber \\%\label{tp2} \\
 & \leq 8\bar{g} \frac{r_0^2}{(1-\beta)^2s^2}+\frac{8}{(1-\beta)^{\alpha}(\alpha-2)}\frac{1}{s^{\alpha}}\frac{(2(1-\beta))^{\alpha-2} s^{\alpha -2} }{{r_0}^{\alpha-2}} \nonumber \\%\label{tp3} \\
 & = \frac{8}{(1-\beta)^2s^2} \left( {\bar{g}r_0^2}+ \frac{2^{\alpha-2}}{(\alpha-2)r_0^{\alpha-2}} \right) \nonumber \\%\label{tp4} \\
 & = \frac{c_2}{s^2} \label{tp5}
\end{align}
where, we set
\begin{equation} 
c_2=\frac{8}{(1-\beta)^2} \left( {\bar{g}r_0^2}+ \frac{2^{\alpha-2}}{(\alpha-2)r_0^{\alpha-2}} \right)
\end{equation}

By \eqref{tp5}, the effective network radius is at most ${c_2}/{(k-1)s^2}$. Therefore, by Theorem \ref{thm1},
\begin{equation}
\EbNomin \geq \frac{(k-1)s^2 \log_e2}{c_2} \label{tp55}
\end{equation}

For the achievable part, consider the algorithm
\begin{equation}
\text{\texttt{FLOOD}} \left( \frac{N_0 \log_e2}{g(2\sqrt{2}s\l)}+\epsilon_1, \frac{N_0 \log_e2}{ \sum_{\ell =1}^{\l} \ell \, g(2\sqrt{2}s\ell) }+\epsilon_1 \right) \label{case3algorithm}
\end{equation}
for any $\epsilon_1>0$.

Before analyzing its energy consumption, let us first see why should this algorithm work. The maximum distance between two nodes belonging to cells that are $\ell \geq 1$ vertical, horizontal or diagonal steps away is $(\ell+1)\sqrt{2}s\leq 2\sqrt{2}s\ell$. Therefore, transmitting with energy per bit $\left( {N_0\log_e2}/{g(2\sqrt{2}s\l)}\right) +\epsilon_1$ ensures that any node within $\l$ steps can decode the message reliably. This is what the source node does. Hence, by the end of the first time slot, any cell belonging to the set 
\begin{equation}
\mathcal{S}_1\triangleq \{ \cell (x,y): \max \{x,y\} \leq (\l-1)+1 \}
\end{equation}
can decode the message reliably. For any $T< T_k \triangleq \sqrt{k}-\l+1$, define 
\begin{equation}
\mathcal{S}_T \triangleq \{ \cell (x,y): \max \{x,y\} \leq (\l-1)+T \}
\end{equation}
Suppose that by the end of time slot $T < T_k-1$, the set of cells which have decoded the message is a superset of $\mathcal{S}_T$. We claim that for any node in the set of cells $\mathcal{S}_{T+1}\setminus \mathcal{S}_{T}$, the total received energy per bit by the end of time slot $T+1$ due to transmissions from nodes in $\mathcal{S}_{T}$ is greater than $N_0\log_e2$. This is true since for any $\ell \leq \l$, any node in the set of cells $\mathcal{S}_{T+1}\setminus \mathcal{S}_{T}$ has at least $\ell$ distinct nodes in $\mathcal{S}_T$ which are exactly $\ell$ vertical, horizontal or diagonal steps away. Note that these nodes at a step distance of $\ell$ are at most $2\sqrt{2}s\ell$ distance away. Since all the nodes in $\mathcal{S}_T$ have transmitted by the end of time slot $T+1$ with an energy per bit of $\left( {N_0 \log_e2}/{ \sum_{\ell=1}^{\l} \ell\, g(2\sqrt{2}s\ell) } \right) +\epsilon_1$, the total received energy per bit at any node in $\mathcal{S}_{T+1}\setminus \mathcal{S}_T$ is greater than $N_0 \log_e2$. Thus, the nodes in $\mathcal{S}_{T+1}$ are covered and by induction, all the nodes are covered by the end of time slot $T_k$.

Since $\l \leq {r_0}/{((1-\beta)s)}$, for all $\ell\leq \l$
\begin{align}
g(2\sqrt{2}s\ell) & \geq g(2\sqrt{2}s \l) \nonumber \\
    & \geq {\left(2\sqrt{2}\frac{r_0}{(1-\beta)}\right)^{-\alpha}} \nonumber \\
    & = \frac{(1-\beta)^{\alpha}}{8^{\alpha/2}r_0^{\alpha}} \label{equation132.0}
\end{align}

Using \eqref{equation132.0}, the total energy consumption of \eqref{case3algorithm} can be bounded by
\begin{align}
{E_b}_{\text{total}} & \leq \frac{N_0 \log_e2}{g(2\sqrt{2}s \l)}+\frac{(k-1)N_0\log_e2}{\sum_{\ell =1}^{\l}\ell\, g(2\sqrt{2}s\ell) }+k \epsilon_1 \nonumber \\
 & \leq \frac{8^{\alpha/2}r_0^{\alpha}N_0 \log_e2}{(1-\beta)^{\alpha}}\left( 1+ \frac{2(k-1)}{\l (\l+1)} \right)+k \epsilon_1 \nonumber \\
 & \leq \frac{8^{\alpha/2}r_0^{\alpha}N_0 \log_e2}{(1-\beta)^{\alpha}}\left( 1+ \frac{4(1-\beta)^2s^2(k-1)}{r_0^2} \right) +k \epsilon_1 \label{117}
\end{align}
for any $\epsilon_1>0$. As before, taking the infimum of all the ${E_b}_{\text{total}}$ values removes the $+k\epsilon_1$ term from the right hand side of \eqref{117} to yield a bound on ${E_b}_{\text{flood}}$.

Therefore, from \eqref{tp55} and \eqref{117}, the upper and lower bounds are related by
\begin{align}
 {E_b}_{\text{flood}} & \leq \frac{8^{\alpha/2}r_0^{\alpha}c_2}{(1-\beta)^{\alpha}}\left( \frac{1}{s^2(k-1)}+ \frac{4(1-\beta)^2}{r_0^2} \right)\, {E_b}_{\min} \nonumber \\
& \leq \frac{8^{\alpha/2}r_0^{\alpha}c_2}{(1-\beta)^{\alpha}}\left( \frac{1}{r_0^2}+ \frac{4(1-\beta)^2}{r_0^2} \right)\, {E_b}_{\min} \label{tp6} \\
& = \frac{8^{\frac{\alpha}{2}+1}r_0^{\alpha}}{(1-\beta)^{\alpha+2}}\left( \bar{g}+\frac{2^{\alpha-2}}{(\alpha-2)r_0^{\alpha}} \right)  \left( 1+ 4(1-\beta)^2 \right)\cdot {E_b}_{\min} \label{tp7}
\end{align}
Note that we have used the condition $(k-1)s^2\geq r_0^2$ in \eqref{tp6}. 

Finally, putting together the results of all three cases, we get the statement of Theorem \ref{RDN}.
\end{IEEEproof}

\section{Discussion and Conclusion}
\label{conclusion}

We established upper and lower bounds on the minimum energy requirement for communicating a message to all the nodes in the network, for various classes of wireless networks. Theorem \ref{thm1} establishes a lower bound on the energy requirement for multicasting in a general static wireless network when the channel state information is not available at the transmitters. This lower bound is based on a fundamental quantity --- effective network radius, which depends only on the expected power gains between pairs of nodes. While the lower bounds based on effective network radius can be arbitrarily weak for certain networks, they are (nearly) order-optimal for at least two classes of large random networks. The near optimality is demonstrated by constructing a multi-stage decode and forward flooding algorithm which requires very little information at the nodes. Therefore, as far as the order of scaling is concerned, knowledge about the node locations or the channel conditions (at the receivers) does not buy much in large random networks.

For dense random networks, the minimum energy per bit is shown to grow linearly with the area $\Ak$ of the network. Compare this with a \emph{single shot transmission} scheme which tries to broadcast the message in a single transmission by the source and, thus, requires energy proportional to $\Ak^{\alpha/2}$. Furthermore, the proposed flooding algorithm is simple, decentralized, and does not require any channel state or node location information. However, the single shot transmission has the advantage of a small delay (just a single time slot) which does not grow with the network size. Also, the single shot transmission can deal with inhomogeneity in the node distribution.

For extended random networks, the lower bound on the minimum energy per bit is shown to be proportional to the number of nodes. However, in extended networks, deviation of the node placement from a `regular' placement has a more detrimental effect on the energy consumption: a factor of at most $(\log k)^{\alpha/2}$ above the converse bound.

In both cases of random networks, the bounds are probabilistic but are violated by at most a finite number of node cardinalities with probability 1. An asymptotic analysis is essential to drive the probability of deviation from a `regular' distribution of nodes to zero. On the other hand, if there is regularity in node placement, we can study the finite case. For finite regular networks, it is shown that not only are the upper and lower bounds valid for all network sizes, but these bounds are within a constant factor of each other for all density regimes as well.

Observe also that the proposed flooding algorithm is inherently fair in the sense that all the non-source nodes expend the same amount of energy. Moreover, in all the instances of the algorithm studied here, the source node spends at least as much energy as each of the other nodes.

\appendices

\section{Proof of Lemma \ref{lemmaconverse}}
\label{prooflemmaconverse}
\begin{IEEEproof}
Consider a $(n,M,E_{\text{total}},\epsilon)$ code over $n$ channel uses. Let $E_{i,t}$ be the expected energy consumption of node $i$ at time $t$, for $i=1,...,k$ and $t=1,...,n$. Therefore, the total energy consumption at node $i$ is 
\begin{equation}
E_i=\sum_{t=1}^nE_{i,t}
\end{equation} 
which satisfy
\begin{equation}
\sum_{i=1}^kE_i=E_{\text{total}} \label{119.5}
\end{equation} 

For the rest of the proof, $\mathbf{x}^{(n)}=(x_1^{(n)},...,x_k^{(n)})$ denotes the set of transmissions at all the nodes. $\mathbf{x}_t=(x_{1,t},x_{2,t},...,x_{k,t})$ is the set of symbols transmitted by all the nodes at time $t$. %We use the notation $\mathbf{h}_{j,t}=(h_{1j,t},...,h_{(j-1)j,t},0,h_{(j+1)j,t},...,h_{kj,t})^T$ to denote the collection of all the channel coefficients to node $j$ at time $t$.

For each node $j\in \Rset$, we derive a form of \emph{cut-set} or \emph{max-flow min-cut} bound (see \cite[Theorem 4]{CoverGamalRelay} \cite[Theorem 15.10.1]{CoverThomas}) in the following steps \eqref{ser1}--\eqref{conend}. 

\begin{align}
(1-\epsilon)\: {\log_2 M} & \leq I(x_1^{(n)};y_{j}^{(n)})+1 \label{ser1} \\
 & \leq I(\mathbf{x}^{(n)};y_{j}^{(n)})+1 \label{ser2} \\
 & = \sum_{t=1}^{n} I(\mathbf{x}^{(n)};y_{j,t}|y_{j,1},...,y_{j,t-1})+1 \label{ser3} \\
 & = \sum_{t=1}^{n} I(\mathbf{x}_{t};y_{j,t}|y_{j,1},...,y_{j,t-1})+1 \label{ser4} \\
 & \leq \sum_{t=1}^{n} I(\mathbf{x}_{t};y_{j,t})+1 \label{ser5} \\
 & \leq \sum_{t=1}^{n} I(\mathbf{x}_{t};y_{j,t}|\mathbf{h}_{j,t})+1 \label{ser5.5} \\
 & \leq \sum_{t=1}^{n} \sup_{\substack{P_{\mathbf{x}_t}:\\ \E[|x_{i,t}|^2]\leq E_{i,t} \text{ for }i=1,...,k }} I(\mathbf{x}_{t};y_{j,t}|\mathbf{h}_{j,t})+1 \label{ser6} \\
 & \leq  n \, \sup_{\substack{P_\mathbf{x}:\\ \E[|x_{i}|^2]\leq E_{i}/n}} I(\mathbf{x};y_{j}|\mathbf{h}_{j})+1 \label{conend}
\end{align}
where \eqref{ser1} is due to Fano's inequality. Inequality \eqref{ser2} is by expanding the set of random variables to $\mathbf{x}^{(n)}$. Applying the chain rule for mutual information to \eqref{ser2} gives us \eqref{ser3}. Step \eqref{ser4} follows from the fact that $y_{j,t}$ depends on $\mathbf{x}^{(n)}$ only through the current transmissions $\mathbf{x}_t$. Similarly, since $(y_{j,1},...,y_{j,t-1})$---$\mathbf{x}_t$---$y_{j,t}$ form a Markov chain, \eqref{ser5} is also true. The random variables $\mathbf{x}_t$ and $\mathbf{h}_{j,t}=(h_{1j,t},...,h_{(j-1)j,t},0,h_{(j+1)j,t},...,h_{kj,t})^T$ are independent of each other, justifying \eqref{ser5.5}. Applying the energy restriction $E_{i,t}$ on the symbol $x_{i,t}$ gives us \eqref{ser6}. Finally, \eqref{conend} is due to the concavity of the mutual information in the cost (in this case, power). Observe that if the supremum of the mutual information in the right hand side of \eqref{conend} is zero, then the mutual information term in the right hand side of \eqref{ser1} is also zero which means that no reliable communication (i.e., $\epsilon \rightarrow 0$) is possible for large message sets ($\log_2 M >1$).

Since inequality \eqref{conend} is valid for all nodes $j \in \Rset$, we can write
\begin{equation}
(1-\epsilon)\: {\log_2 M} \leq \min_{j\in \Rset}\; n \, \sup_{\substack{P_\mathbf{x}:\\ \E[|x_{i}|^2]\leq E_{i}/n}} I(\mathbf{x};y_{j}|\mathbf{h}_{j})+1 \label{eqnumber119}
\end{equation}

Therefore, the energy per bit of the code is 
\begin{align}
\frac{E_{\text{total}}}{\log_2 M} & \geq \frac{(1-\epsilon)E_{\text{total}}}{n \, \min_{j\in \Rset} \sup_{\substack{P_\mathbf{x}:\\ \E[|x_{i}|^2]\leq E_{i}/n}} I(\mathbf{x};y_{j}|\mathbf{h}_{j})+1} \\
 & = (1-\epsilon) \bigdiv \left( \frac{\min_{j\in \Rset} \sup_{\substack{P_\mathbf{x}:\\ \E[|x_{i}|^2]\leq E_{i}/n}} I(\mathbf{x};y_{j}|\mathbf{h}_{j})}{\sum_{i=1}^kE_i/n}+\frac{1}{E_{\text{total}}} \right) \label{equationnumber147} \\
& \geq (1-\epsilon)\bigdiv \left( \sup_{\substack{P_1,P_2,...,P_k\geq 0, \\ \sum_{i=1}^kP_i>0 }} \min_{j\in \Rset} \frac{\sup_{\substack{P_\mathbf{x}:\\ \E[|x_{i}|^2]\leq P_i}} I(\mathbf{x};y_{j}|\mathbf{h}_{j})}{\sum_{i=1}^kP_i} +\frac{1}{E_{\text{total}}} \right) \label{equationnumber122}
\end{align}
where to get \eqref{equationnumber122}, we have substituted $E_{i}/n$ by $P_i$ in \eqref{equationnumber147} and taken supremum over $P_i$ for all $i=1,...,k$. 
Furthermore, if the first term in the denominator of \eqref{equationnumber122} is zero, then reliable communication is not possible at any finite transmission power. 

Recall that if $E_b$ is $\epsilon$-achievable, then for all $\delta>0$, there is an $E_0\in \mathbb{R}_+$ such that for every $E_{\text{total}}\geq E_0$,
\begin{equation}
\frac{E_{\text{total}}}{\log_2 M} < E_b + \delta
\end{equation}

This implies that
\begin{equation}
\lim_{E_{\text{total}}\rightarrow \infty} \sup\quad  (1-\epsilon)\bigdiv \left( \sup_{\substack{P_1,...,P_k\geq 0,\\ \sum_{i=1}^kP_i>0}} \min_{j\in \Rset} \frac{\sup_{\substack{P_\mathbf{x}:\\ \E[|x_{i}|^2]\leq P_i}} I(\mathbf{x};y_{j}|\mathbf{h}_j)}{\sum_{i=1}^kP_i} +\frac{1}{E_{\text{total}}} \right) \leq E_b \label{132}
\end{equation}

Moreover, if $E_b$ is an achievable energy per bit value, then we can supremize the left side of \eqref{132} for all $0< \epsilon <1$ to get
\begin{equation}
E_b \geq \inf_{\substack{P_1,...,P_k\geq 0,\\ \sum_{i=1}^kP_i>0}} \max_{j\in \Rset} \frac{\sum_{i=1}^k P_i}{\sup_{\substack{P_\mathbf{x}:\\ E[|x_{i}|^2]\leq P_i}} I(\mathbf{x};y_{j}|\mathbf{h}_j)} \label{lastconvst}
\end{equation}

Therefore, ${E_b}_{\min}$ should also satisfy \eqref{lastconvst}.

\end{IEEEproof}

\section{Proof of Lemma \ref{gooddist}}
\label{proofgooddist}
\begin{IEEEproof} 

First, note that
\begin{align}
\P[\mathcal{D}_k^c] & =  \P[\exists \cell \in \mathcal{C}:\k(\cell)\notin \left[(1-\delta)\frac{(k-1)s^2}{\Ak},(1+\delta)\frac{(k-1)s^2}{\Ak}+1\right) ] \nonumber \\
 & \leq \P[\exists \cell \in \mathcal{C}:\k(\cell) < (1-\delta)\frac{(k-1)s^2}{\Ak}] \nonumber \\
 & \quad + \P[\exists \cell \in \mathcal{C}:\k(\cell)\geq (1+\delta)\frac{(k-1)s^2}{\Ak}+1 ] \label{34p5}
\end{align}
by taking the union bound over the two ways of violating the condition for $\mathcal{D}_k$. Each of these two terms can be further union bounded over all the cells. We note that all the cells are identical as far as number of nodes in them is concerned, except for the fact that the origin cell, say $\cell_1$, already contains an additional node (the source node). Therefore, the origin cell will have the maximum probability of having greater than $((1+\delta){(k-1)s^2}/{\Ak})+1$ nodes and the non-origin cells will have the maximum probability of having less than $(1-\delta){(k-1)s^2}/{\Ak}$ nodes. Let $\cell_2$ be a representative cell for non-origin cells. Noting that there are $\Ak/s^2$ cells, we have the following union bound on \eqref{34p5}.
%\begin{align}
%\P[\mathcal{D}_k^c] & \leq \frac{\Ak}{s^2} \left( P[k(\cell_1)\geq (1+\delta)\frac{(k-1)s^2}{\Ak}+1 ] \right. \nonumber \\
% & \qquad \left. + P[k(\cell_2)< (1-\delta)\frac{(k-1)s^2}{\Ak}] \right) \label{832}
%\end{align}
\begin{equation}
\P[\mathcal{D}_k^c] \leq \frac{\Ak}{s^2} \left( P[\k(\cell_1)\geq (1+\delta)\frac{(k-1)s^2}{\Ak}+1 ] + P[\k(\cell_2)< (1-\delta)\frac{(k-1)s^2}{\Ak}] \right) \label{832}
\end{equation}
To further bound the probabilities in \eqref{832}, we use the Chernoff bound. Let $X_i$ be the indicator function of node $i$ falling in $\cell_2$. Then, for $i=2,...,k$
\begin{displaymath}
X_i = \left\{ \begin{array}{ll}
0 & \text{w.p. } 1-\frac{s^2}{\Ak} \\
1 & \text{w.p. } \frac{s^2}{\Ak}
\end{array} \right. 
\end{displaymath}
are independent random variables. The probability that cell $\cell_2$ contains fewer than $(1-\delta){(k-1)s^2}/{\Ak}$ nodes is equivalent to evaluating  
\begin{equation}
\P\left[ \sum_{i=2}^{k}X_i<(1-\delta)\frac{(k-1)s^2}{\Ak} \right] \label{35.5}
\end{equation}
By a simple change of variables $X'_i=X_i-({s^2}/{\Ak})$, we get that 
\begin{displaymath}
X'_i = \left\{ \begin{array}{ll}
-\frac{s^2}{\Ak} & \text{w.p. } 1-\frac{s^2}{\Ak} \\
1-\frac{s^2}{\Ak} & \text{w.p. } \frac{s^2}{\Ak}
\end{array} \right. \qquad i=2,...,k
\end{displaymath}
which satisfies \cite[Assumptions A.1.3]{Alon}. So, by the Chernoff bound \cite[Theorem A.1.13]{Alon},
\begin{equation}
\P\left[\sum_{i=2}^{k}X'_i<(-\delta)\frac{(k-1)s^2}{\Ak}\right] <  \exp \left(-\delta^2\frac{(k-1)s^2}{2\Ak}\right) \label{upper1}
\end{equation}
The right hand side of \eqref{upper1} provides an upper bound on \eqref{35.5} and thus an upper bound on the probability of $\k(\cell_2)$ being less than $(1-\delta){(k-1)s^2}/{\Ak}$.

Retaining the variable $X'$ and applying \cite[Theorem A.1.11]{Alon}, we also get 
\begin{align}
\P[\k(\cell_1)\geq (1+\delta)\frac{(k-1)s^2}{\Ak}+1] & = \P\left[ \sum_{i=2}^{k}X'_i \geq \delta\, \frac{(k-1)s^2}{\Ak}\right] \nonumber \\
 & <  \exp \left(-\delta^2(1-\delta)\frac{(k-1)s^2}{2\Ak} \right) \label{lower2}
\end{align}

Using \eqref{upper1} and \eqref{lower2} in \eqref{832} and noticing that the right hand side of \eqref{lower2} is greater than the right hand side of \eqref{upper1}, we get the final result.
\end{IEEEproof}

\section{Proof of Lemma \ref{lemma3p0}}
\label{prooflemma3p0}
\begin{IEEEproof}

Label the $k$ cells as $\cell_1,\cell_2,...,\cell_{k}$, where $\cell_1$ is the origin cell. Let $X_i$ be a random variable indicating the cell in which the non-source node $i$ falls. We will also use $X_i$ to indicate whether the node $i$ falls in the window portion or the non-window portion of the cell, in the following manner. Let $X_i$ take the integer values from $-k$ to $+k$ excluding zero. If $X_i=+j$ for some $j\in \left\{1,...,k \right\}$, it implies that the node $i$ falls into the windowed portion (with the area $\beta^2/\lambda$) of $\cell_j$. If $X_i=-j$ for some $j\in \left\{1,...,k \right\}$, it implies that the node $i$ falls into the non-windowed portion (with the area $(1-\beta^2)/\lambda$) of $\cell_j$. Clearly, the $X_i$s are i.i.d. with common distribution $X$ given by
\begin{equation}
X = \left\{ \begin{array}{ll}
+j & \text{w.p. } \frac{\beta^2}{k} \\
-j & \text{w.p. } \frac{1-\beta^2}{k}
\end{array} \right. \text{ for all }j\in \left\{1,...,k \right\} \label{eq82.5}
\end{equation}

Though all the other cells are identical, $\cell_1$ already contains the source node outside its window, so for the next set of calculations, we will only be dealing with the $k-1$ non-source nodes and $k-1$ non-origin cells. 

Consider the function $f: \left\{ -k,...,-1,+1,...,k \right\} ^ {k-1} \mapsto \mathbb{N}$, that counts the number of good non-origin cells in a realization of $(X_2,X_3,...,X_{k})$, i.e.,
\begin{equation}
f(x_2,x_3,...,x_k)=\sum_{i=2}^{k} \mathbf{1}\left\{ \exists j \in \{2,...,k\}: x_j=+i, \text{ and }x_{\ell} \neq \pm i\, \forall \ell \in \{2,...,j-1,j+1,...,k\}  \right\}
\end{equation}

Taking expectation of the function $f$, we get
\begin{align}
\E[f(X_2,...,X_k)] & = \sum_{i=2}^{k} \E \left[\mathbf{1} \left\{ \exists j \in \{2,...,k\}: X_j=+i, \text{ and }X_{\ell} \neq \pm i\, \forall \ell \in \{2,...,j-1,j+1,...,k\}  \right\} \right] \nonumber \\
 & = \left( k-1 \right) \P\left[\exists j \in \{2,...,k\}: X_j=+2, \text{ and }X_{\ell} \neq \pm 2\, \forall \ell \in \{2,...,j-1,j+1,...,k\} \right] \label{90p5}
\end{align}
where \eqref{90p5} follows by observing that all the non-origin cells are identical; so, in \eqref{90p5}, we pick cell 2 as a representative non-origin cell. Using \eqref{eq82.5} we can evaluate the probability in \eqref{90p5} which is the probability that $X_j$ is $+2$ for exactly one $j\in \{2,...,k\}$ (which can happen in $k-1$ ways) and the rest $k-2$ nodes fall outside the cell 2. Therefore,
\begin{align}
\E[f(X_2,...,X_k)] %& = \left( \frac{\Ak}{s^2}-1 \right) (k-1) \frac{\beta^2 s^2}{\Ak}\left(1-\frac{s^2}{\Ak} \right)^{k-2} \nonumber \\
 & = \beta^2 \frac{(k-1)^k}{k^{k-1}} \label{meanoffunction}
\end{align}
%where the first step uses the fact that all the non-origin cells are identical, and the second step is by evaluating the probability using \eqref{eq82.5}.

Our task is to bound the deviation of the function $f$ from its mean value.
We do this using McDiarmid's inequality, which is a generalization of the Chernoff bound:
\begin{lemma}[McDiarmid's Inequality \cite{McDiarmid}]
\label{MI}
Let $X_1,X_2,...,X_n$ be independent random variables, with $X_i$ taking values in a set $\mathcal{X}$ for each $i$. Suppose that the function $f:\mathcal{X}^n\mapsto \mathbb{R}$ satisfies
\begin{equation}
|f(\mathbf{x})-f(\mathbf{x}')| \leq c_i
\end{equation}
whenever $\mathbf{x}$ and $\mathbf{x}'$ differ only in the $i^{th}$ coordinate. Then, for any $t>0$,
\begin{equation}
\P[|f(X_1,X_2,...,X_n)-\E[f(X_1,X_2,...,X_n)]|\geq t] \leq 2\exp\left({-\frac{2t^2}{\sum_{i=1}^{n}c_{i}^2}}\right)
\end{equation}
\end{lemma}

In our case, the value of function $f$ varies by at most 2 whenever only one component changes. This maximal variation corresponds to those situations when a node belonging to a good cell moves to another good cell, thus destroying the `goodness' of both of them. Similar situation can be imagined when two good cells are generated by movement of one node. Therefore, 
\begin{equation}
c_i \leq 2 \quad \text{ for all }i=2,3,...,k
\end{equation}

Setting $t=\delta \beta^2 {(k-1)^k}/{k^{k-1}}$ and using the value of $\E[f(X_2,...,X_k)]$ from \eqref{meanoffunction} in Lemma \ref{MI}, we immediately get \eqref{appmcdiarmid}.

\end{IEEEproof}

\bibliographystyle{IEEEtranS}  
\bibliography{IEEEabrv,sensornetworks}

% Generated by IEEEtranS.bst, version: 1.12 (2007/01/11)
\begin{thebibliography}{10}
\providecommand{\url}[1]{#1}
\csname url@samestyle\endcsname
\providecommand{\newblock}{\relax}
\providecommand{\bibinfo}[2]{#2}
\providecommand{\BIBentrySTDinterwordspacing}{\spaceskip=0pt\relax}
\providecommand{\BIBentryALTinterwordstretchfactor}{4}
\providecommand{\BIBentryALTinterwordspacing}{\spaceskip=\fontdimen2\font plus
\BIBentryALTinterwordstretchfactor\fontdimen3\font minus
  \fontdimen4\font\relax}
\providecommand{\BIBforeignlanguage}[2]{{%
\expandafter\ifx\csname l@#1\endcsname\relax
\typeout{** WARNING: IEEEtranS.bst: No hyphenation pattern has been}%
\typeout{** loaded for the language `#1'. Using the pattern for}%
\typeout{** the default language instead.}%
\else
\language=\csname l@#1\endcsname
\fi
#2}}
\providecommand{\BIBdecl}{\relax}
\BIBdecl

\bibitem{Sensor}
I.~Akyildiz, W.~Su, Y.~Sankarasubramaniam, and E.~Cayirci, ``A survey on sensor
  networks,'' \emph{{IEEE} Commun. Mag.}, vol.~40, no.~8, pp. 102--114, Aug.
  2002.

\bibitem{Alon}
N.~Alon and J.~Spencer, \emph{The Probabilistic Method}, 2nd~ed.\hskip 1em plus
  0.5em minus 0.4em\relax {New York}, {NY}, {USA}: John Wiley \& Sons, 2000.

\bibitem{CYG}
X.~Cai, Y.~Yao, and G.~Giannakis, ``Achievable rates in low-power relay links
  over fading channels,'' \emph{{IEEE} Trans. Commun.}, vol.~53, no.~1, Jan.
  2005.

\bibitem{Caire}
G.~Caire, D.~Tuninetti, and S.~Verd{\'u}, ``Suboptimality of {TDMA} in the
  low-power regime,'' \emph{{IEEE} Trans. Inf. Theory}, vol.~50, no.~4, pp.
  608--620, Apr. 2004.

\bibitem{CoverGamalRelay}
T.~Cover and A.~E. Gamal, ``Capacity theorems for the relay channel,''
  \emph{{IEEE} Trans. Inf. Theory}, vol.~25, no.~5, pp. 572--584, Sep. 1979.

\bibitem{CoverThomas}
T.~Cover and J.~Thomas, \emph{Elements of information theory}, 2nd~ed.\hskip
  1em plus 0.5em minus 0.4em\relax {New York}, {NY}, {USA}: John Wiley \& Sons,
  2006.

\bibitem{Dana}
A.~F. Dana and B.~Hassibi, ``On the power efficiency of sensory and ad hoc
  wireless networks,'' \emph{{IEEE} Trans. Inf. Theory}, vol.~52, no.~7, pp.
  2890--2914, Jul. 2006.

\bibitem{Mammen}
A.~{El Gamal} and J.~Mammen, ``Optimal hopping in ad hoc wireless networks,''
  in \emph{Proc. 25th Annual Joint Conference of the {IEEE} Computer and
  Communications Societies ({INFOCOM})}, 2006, pp. 1--10.

\bibitem{Zahedi}
A.~{El Gamal}, M.~Mohseni, and S.~Zahedi, ``Bounds on capacity and minimum
  energy-per-bit for {AWGN} relay channels,'' \emph{{IEEE} Trans. Inf. Theory},
  vol.~52, no.~4, pp. 1545--1561, Apr. 2006.

\bibitem{Riedi}
A.~K. Haddad, V.~Ribeiro, and R.~Riedi, ``Broadcast capacity in multihop
  wireless networks,'' in \emph{Proc. 12th Annual International Conference on
  Mobile Computing and Networking ({MOBICOM})}, 2006, pp. 239--250.

\bibitem{ScaglioneOLA2}
Y.~Hong and A.~Scaglione, ``Energy-efficient broadcasting with cooperative
  transmissions in wireless sensor networks,'' \emph{{IEEE} Trans. Wireless
  Commun.}, vol.~5, no.~10, pp. 2844--2855, Oct. 2006.

\bibitem{Wornell}
J.~Laneman, D.~Tse, and G.~Wornell, ``Cooperative diversity in wireless
  networks: Efficient protocols and outage behavior,'' \emph{{IEEE} Trans. Inf.
  Theory}, vol.~50, no.~12, pp. 3062--3080, Dec. 2004.

\bibitem{Lapidoth}
A.~Lapidoth, I.~Telatar, and R.~Urbanke, ``On wide-band broadcast channels,''
  \emph{{IEEE} Trans. Inf. Theory}, vol.~49, no.~12, pp. 3250--3258, Dec. 2003.

\bibitem{flood}
H.~Lim and C.~Kim, ``Multicast tree construction and flooding in wireless ad
  hoc networks,'' in \emph{Proc. 3rd {ACM} International Workshop on Modeling,
  Analysis and Simulation of Wireless and Mobile Systems ({MSWIM})}, 2000, pp.
  61--68.

\bibitem{Maric}
I.~Maric and R.~D. Yates, ``Cooperative multihop broadcast for wireless
  networks,'' \emph{{IEEE} J. Sel. Areas Commun.}, vol.~22, no.~6, pp.
  1080--1088, Aug. 2004.

\bibitem{McDiarmid}
C.~McDiarmid, ``On the method of bounded differences,'' \emph{Surveys in
  Combinatorics}, vol. 141, pp. 148--188, 1989.

\bibitem{MergenGastpar2}
B.~S. Mergen and M.~Gastpar, ``The scaling of the broadcast capacity of
  extended wireless networks with cooperative relays,'' in \emph{Proc. 41st
  {A}silomar Conference on Signals, Systems and Computers ({ACSSC})}, 2007, pp.
  873--877.

\bibitem{MergenGastpar1}
------, ``On the broadcast capacity of wireless networks with cooperative
  relays,'' \emph{{IEEE} Trans. Inf. Theory}, submitted for publication.

\bibitem{ScaglioneJSAC}
B.~S. Mergen and A.~Scaglione, ``On the power efficiency of cooperative
  broadcast in dense wireless networks,'' \emph{{IEEE} J. Sel. Areas Commun.},
  vol.~25, no.~2, pp. 497--507, Feb. 2007.

\bibitem{ScaglioneITTrans}
B.~S. Mergen, A.~Scaglione, and G.~Mergen, ``Asymptotic analysis of multistage
  cooperative broadcast in wireless networks,'' \emph{{IEEE} Trans. Inf.
  Theory}, vol.~52, no.~6, pp. 2531--2550, Jun. 2006.

\bibitem{Obraczka}
K.~Obraczka, K.~Viswanath, and G.~Tsudik, ``Flooding for reliable multicast in
  multi-hop ad hoc networks,'' \emph{Wireless Networks}, vol.~7, no.~6, pp.
  627--634, Nov. 2001.

\bibitem{ScaglioneOLA1}
A.~Scaglione and Y.~Hong, ``Opportunistic large arrays: cooperative
  transmission in wireless multihop ad hoc networks to reach far distances,''
  \emph{{IEEE} Trans. Signal Process.}, vol.~51, no.~8, pp. 2082--2092, Aug.
  2003.

\bibitem{Sendonaris}
A.~Sendonaris, E.~Erkip, and B.~Aazhang, ``User cooperation diversity - part
  {I}: System description,'' \emph{{IEEE} Trans. Commun.}, vol.~51, no.~11, pp.
  1927--1938, Nov. 2003.

\bibitem{Sendonaris2}
------, ``User cooperation diversity - part {II}: Implementation aspects and
  performance analysis,'' \emph{{IEEE} Trans. Commun.}, vol.~51, no.~11, pp.
  1939--1948, Nov. 2003.

\bibitem{Spectral}
S.~Verd{\'u}, ``Spectral efficiency in the wideband regime,'' \emph{{IEEE}
  Trans. Inf. Theory}, vol.~48, no.~6, pp. 1319--1343, Jun. 2002.

\bibitem{UnitCost}
------, ``On channel capacity per unit cost,'' \emph{{IEEE} Trans. Inf.
  Theory}, vol.~36, no.~5, pp. 1019--1030, Sep. 1990.

\bibitem{BT}
J.~E. Wieselthier, G.~D. Nguyen, and A.~Ephremides, ``On the construction of
  energy-efficient broadcast and multicast trees in wireless networks,'' in
  \emph{Proc. 19th Annual Joint Conference of the {IEEE} Computer and
  Communications Societies ({INFOCOM})}, vol.~2, 2000, pp. 585--594.

\bibitem{Camp}
B.~Williams and T.~Camp, ``Comparison of broadcasting techniques for mobile ad
  hoc networks,'' in \emph{Proc. 3rd {ACM} International Symposium on Mobile Ad
  Hoc Networking and Computing ({MOBIHOC 2002})}, 2002, pp. 194--205.

\bibitem{YCG}
Y.~Yao, X.~Cai, and G.~B. Giannakis, ``On energy efficiency and optimum
  resource allocation in wireless relay transmissions,'' \emph{{IEEE} Trans.
  Wireless Commun.}, vol.~4, no.~6, pp. 2917--2927, Nov. 2005.

\bibitem{Zheng}
R.~Zheng, ``Information dissemination in power-constrained wireless networks,''
  in \emph{Proc. 25th Annual Joint Conference of the {IEEE} Computer and
  Communications Societies ({INFOCOM})}, 2006, pp. 1--10.

\end{thebibliography}

\end{document}